\documentclass[a4paper,fleqn,usenatbib]{mnras}

\usepackage{newtxtext,newtxmath}

\usepackage[T1]{fontenc}
\usepackage{ae,aecompl}

\usepackage{graphicx}	
\usepackage{amsmath}	

\title[Dust in discs near massive stars]{Warm millimetre dust in protoplanetary discs near massive stars}
\author[T. J. Haworth]
{\parbox{\textwidth}{Thomas J. Haworth$^{1}$\thanks{E-mail: \texttt{t.haworth@qmul.ac.uk}}
}\vspace{0.4cm}\\
\parbox{\textwidth}{$^{1}$ Astronomy Unit, School of Physics and Astronomy, Queen Mary University of London, London E1 4NS, UK \\
}}

\begin{document}

\date{Accepted ???. Received ???; in original form ???}

\pagerange{\pageref{firstpage}--\pageref{lastpage}} \pubyear{2020}

\maketitle
\label{firstpage}

\begin{abstract}
Dust plays a key role in the formation of planets and its emission also provides one of our most accessible views of protoplanetary discs. {If set by radiative equilibrium with the central star, the temperature of dust in the disc plateaus at around $10-20$\,K in the outer regions.} {However sufficiently nearby massive stars can  heat the outer disc to substantially higher temperatures. In this paper we study the radiative equilibrium temperature of discs in the presence of massive external sources and gauge the effect that it has on millimetre dust mass estimates.} Since millimetre grains are not entrained in any wind we focus on geometrically simple 2D-axisymmetric disc models using radiative transfer calculations with both the host star and an external source. {Recent surveys have searched for evidence of massive stars influencing disc evolution using disc properties as a function of projected separation}.  In assuming a disc temperature of $20$\,K for a disc a distance $D$ from a strong radiation source, disc masses are overestimated by a factor that scales with $D^{-1/2}$ {interior to the separation that external heating becomes important}. This could significantly alter dust mass estimates of discs in close proximity to $\theta^1$C in the Orion Nebular Cluster.  We also make an initial assessment of the effect upon snow lines. Within a parsec of an O star like $\theta^1$C a CO snow line no longer exists, though the water snow line is virtually unaffected except for very close separations of $\leq0.01\,$pc. 
\end{abstract}


\section{introduction}
Although we have known about the impact of environment on circumstellar discs essentially for as long as we have been able to directly image them \citep{1994ApJ...436..194O}, there has recently been a resurgence of interest in the topic. 

{Most stars form in clustered environments \citep[e.g.][]{2003ARA&A..41...57L, 2019ARA&A..57..227K}}. The three main ways that a cluster environment affects discs is through external photoevaporation, dynamical (gravitational) encounters and through compositional inheritence and enrichment \citep[e.g. of short lived radionuclides like Aluminium 26,][though we do not focus on this here]{2019NatAs...3..307L, 2020A&A...644L...1R}. Recent developments with regard to the first two points include a better understanding {of} the effect of dynamics and external photoevaporation in different types of stellar cluster \citep[][]{2001MNRAS.325..449S, 2018MNRAS.478.2700W, 2019MNRAS.490.5678C, 2019MNRAS.485.4893N} which were facilitated by improvements in external photoevaporation models \citep{2018MNRAS.481..452H, 2019MNRAS.485.3895H}, observations of interacting systems \citep[e.g.][]{2018ApJ...859..150R, 2018ApJ...869L..44K} and an ever increasing catalogue of photoevaporating discs \citep[e.g.][]{1999AJ....118.2350H, 2014ApJ...784...82M, 2016ApJ...826L..15K, 2021MNRAS.501.3502H}. The recent linking of possibly distinct exoplanet populations in Gaia phase space overdensities also provides a tantalizing hint of the impact of environment (at the formation stage, or over a longer period of time) on the resulting planets themselves \citep{2020Natur.586..528W, 2020ApJ...905L..18K, 2021arXiv210301974L}. 

Another powerful diagnostic comes from finding systematic statistical variation of disc properties in a region. For example in Hubble Space Telescope observations, disc sizes were found to vary with distance from the O star $\theta^1$C by \cite{2013MNRAS.430.3406T} \citep[see also][]{1998AJ....116..322H, 1999AJ....118.2350H}.  The unprecedented sensitivity and resolution that ALMA provides now also allows us to analyse disc dust mass and radius statistics throughout star forming regions \citep[e.g.][]{2014ApJ...784...82M, 2017AJ....153..240A, 2018ApJ...860...77E, 2020ApJ...894...74B, 2020AJ....160..248A}. In particular, trends in disc properties as a function of projected separation from the strongest UV sources is often interpreted as evidence for external photoevaporation \citep{2014ApJ...784...82M, 2017AJ....153..240A, 2018ApJ...860...77E}. However, the dynamical evolution of the cluster can drastically complicate this picture (Parker et al. in preparation). Nevertheless trends do remain that require an explanation.

The most accessible (i.e. least demanding in terms of observing time) insight into the disc properties with ALMA comes from estimating the dust mass from the measured continuum flux $F_\nu$. This is done by integrating the formal solution to the equation of radiative transfer in the limit of being optically thin, constant absorption and emission coefficient, no background source and dust emitting as a blackbody (Planck function $B_\nu$)
\begin{equation}
    I_\nu = B_\nu\left(1 - \exp(-\tau_\nu)\right) \approx B_\nu\tau_\nu
\end{equation}
over the beam solid angle $\Omega_B$ to convert the emergent intensity $I_\nu$ to a flux $F_\nu$	
\begin{equation}
	    F_\nu \approx B_\nu \tau_\nu \Omega_B = B_\nu \kappa_\nu \rho l \Omega_B
	\end{equation}
{where $\kappa_\nu$ is the opacity and $\rho$  and $l$ are the density and thickness of the emitter along the line of sight}.	
The solid angle is then related to the mass 
\begin{equation}
M_d = \rho l \Omega_B D^2
\end{equation}
hence
\begin{equation}
    M_d = \frac{D^2F_\nu}{\kappa_\nu B_\nu(T)}
    \label{equn:massEstimate}
\end{equation}
is the well known expression used to estimate dust masses from a measured flux $F_\nu$ {\citep{1983QJRAS..24..267H}}. This relies upon an assumed distance $D$, opacity at the frequency of observation $\kappa_\nu$ and dust temperature $T$. {In many prior studies of discs the dust temperature and mass was estimated from SED fitting \citep[e.g.][]{1990AJ.....99..924B, 1994ApJ...420..837A, 2005ApJ...631.1134A, 2007ApJ...671.1800A}. } {However often in ALMA mm continuum surveys this dust temperature is assumed to be 20\,K}, which probably applies well to nearby discs to the Sun which are very extended (hundreds of au in size) and expected to be relatively cold over the bulk of the disc mass. \cite{2017A&A...606A..88T} validated this {assumption} further with UV plane fitting of discs in Lupus, finding no obvious dependence on disc temperature. 

So for nearby discs, the host star heating only affects the very inner regions and it does seem reasonable to estimate the disc mass using a cooler temperature like 20\,K {that well represents the temperature of} the bulk of the disc mass. {However when studying discs in relatively close proximity to massive stars ($\sim1\,$pc in the case of $\theta^1$C, as we will demonstrate here), warming of the outer disc by external sources} could mean that disc dust masses in the denser parts of massive clusters are actually being {overestimated} in surveys when a temperature of 20\,K is assumed for all discs. The implications here could be important for understanding the statistical properties of discs in clusters. Furthermore, this external heating could also affect the locations (or even existence) of snow lines in the disc, as well as the nature of grain evolution and even the ability to introduce pressure bumps, which could all affect the planet formation process {in these high radiation environments}. 

External irradiation has been modelled in strong radiation scenarios looking at the effect on the gas dynamics/proplyds \citep[e.g.][]{1999AJ....118.2350H, 2000ApJ...539..258R, 2001ApJ...561..830G, 2019MNRAS.485.3895H}. However these calculations are computationally expensive, usually focus on the dynamics, and when it comes to observables usually focus on the photodissociation region (PDR) and photoionised gas  diagnostics. {\cite{2013ApJ...766L..23W} compared the temperature and chemistry of an isolated disc with an identical disc system irradiated by a strong external field, corresponding to being within $<0.1\,$pc of an O star. They noted substantial heating and impact upon the chemistry (e.g. the CO snow line disappears)}. More intermediate/low FUV radiation fields have also been included in hydrostatic disc models, but with a view to how it affects factors like ionisation and CO photodesorption \citep[e.g.][]{2013ApJ...772....5C, 2016ApJ...816L..21C} rather than the dust radiative equilibrium temperature. \cite{2017A&A...604A..69C} undertook 1D PDR models of proplyds using the \textsc{meudon} code, however in those they compute the dust temperature in the surface layers and the disc temperature is assumed to be uniformly 19.5\,K. \cite{2002ApJ...578..897R} also took a similar approach semi-analytically, considering the proplyd to be a spherical system with non-spherical external irradiation and internal heating by the host star, but again the focus was on the layers external to the circumstellar disc and their infrared emission. 

\cite{2020MNRAS.492.1279S} studied the dynamical evolution of dust in discs with an external photoevaporative wind. They accounted for the fact that only small grains are entrained away from the disc and into a wind/proplyd envelope, as expected from \cite{2016MNRAS.457.3593F} and observed by \cite{2012ApJ...757...78M}, \cite{2017ApJ...840...55B}. However \cite{2020MNRAS.492.1279S} assumed a temperature profile for the disc that is dominated by the central source. \cite{2018MNRAS.474..886N} included a parameterised heating of the outer disc in planet population synthesis models, finding that heating due to the cluster environment is important for suppressing large populations of cold Jupiters, which are not observed (particularly at low metallicity). Their model included the effect of heating on snow line locations, but the locations themselves in different environments were not discussed.

{Here we focus on understanding how external heating by massive stars affects millimetre continuum mass estimates. Furthermore we aim to understand how this in turn affects the statistical variation of disc properties near massive stars and the implications that has for understanding how massive stars impact disc evolution.}

\section{Simple optically thin radiative equilibrium model}
\label{sec:simpleRadEq}
We take a quick first look at the possible impact of an external source by considering an extension to the classic optically thin radiative equilibrium expression using the bolometric luminosity of the host star and external source, similar to the approach taken by \cite{2013MNRAS.430.3406T} in estimating the dust temperature of ONC proplyds.  

Assume that there is a circumstellar disc that is heated to the grain radiative equilibrium temperature by the central host star (luminosity $L_H)$ and an external source (luminosity $L_{ext}$). Also assume that the disc is optically thin. The heating rate for a grain of radius $r_d$ and albedo $\alpha$ at radial distance $R$ from the host star with the external source at distance $D$ is 
\begin{equation}
    H = A_H4\pi r_d^2 \left(\frac{L_H}{4\pi R^2} + \frac{L_{ext}}{4\pi D^2}\right)(1-\alpha)
\end{equation}
where $A_H$ is the fraction of the spherical grain surface that is irradiated. In the case of a planar radiation field travelling through a medium with zero scattering {$A_H=1/4$} because only one side of the grain is irradiated. However, for a more general radiation field, or a planar field incident upon a scattering medium $A_H=1$. We spend the time clarifying this because Monte Carlo radiative transfer codes \citep{1999A&A...344..282L} generally compute the averaged properties in each volume of a calculation, effectively setting $A_H=1$. We assume that the heating only has a radial dependence on the contribution from the host star and that the external heating is uniform. Assuming both sources are blackbodies the heating term becomes
\begin{equation}
    H = A_H4\pi r_d^2 \sigma \left(\frac{R_H^2}{R^2}T_H^4 + \frac{R_{ext}^2}{D^2}T_{ext}^4\right)(1-\alpha). 
\end{equation}

\begin{figure}
    \centering
    \includegraphics[width=\columnwidth]{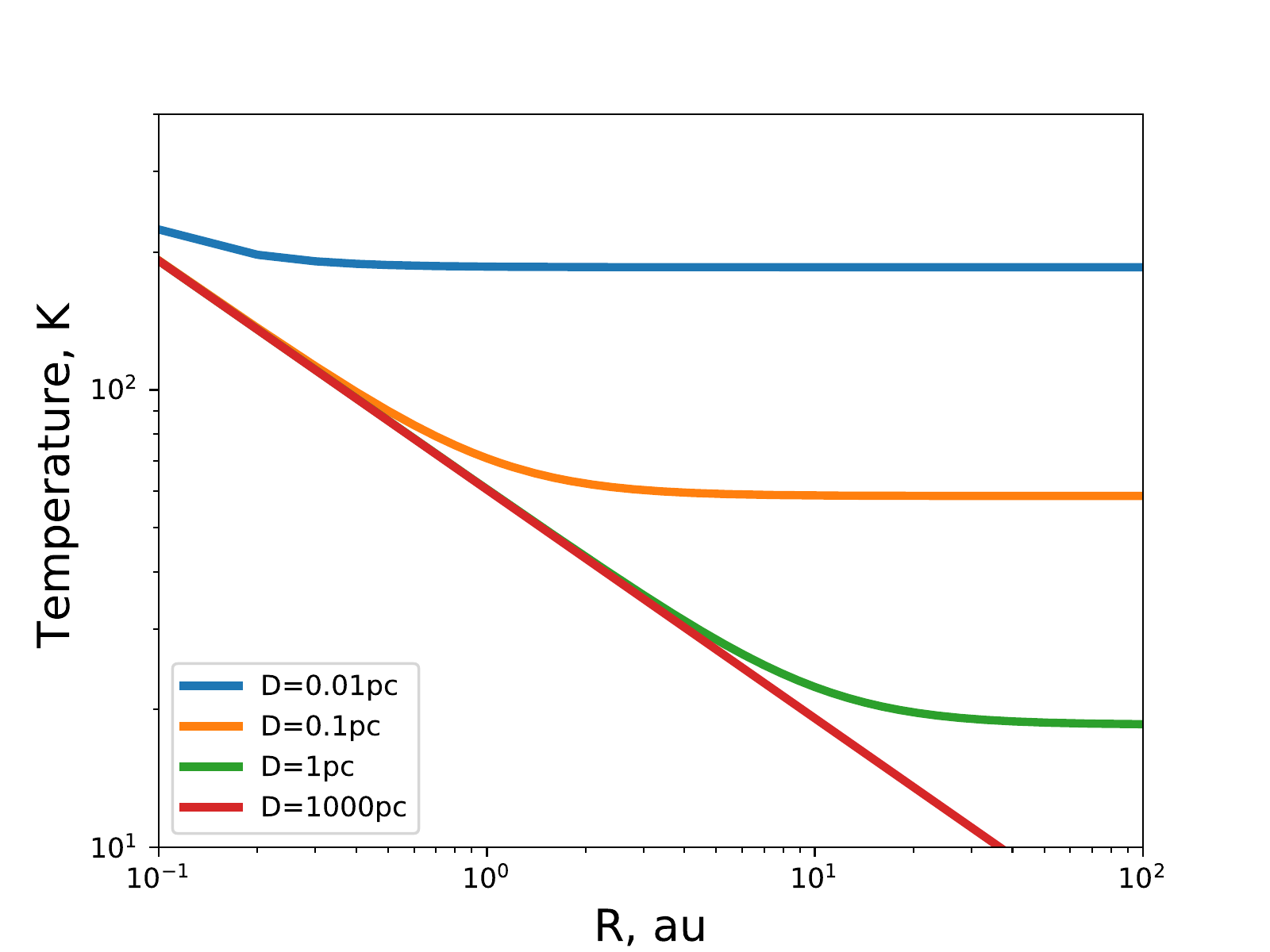}
    \caption{Analytic disc radiative equilibrium radial temperature profiles for a host star with $T_H=2550$, $R_H=0.121$ (Trappist-1) at various distances from a $\theta^1$C analogue. }
    \label{fig:Tprofiles_simple}
\end{figure}

The cooling rate of spherical grains in the disc at temperatre $T_d$ is
\begin{equation}
    C = 4\pi r_d^2 \sigma T_d^4
\end{equation}
Equating the heating and cooling in radiative equilibrium
\begin{equation}
    4\pi r_d^2 \sigma T_d^4 = A_H4\pi r_d^2 \sigma \left(\frac{R_H^2}{R^2}T_H^4 + \frac{R_{ext}^2}{D^2}T_{ext}^4\right)\left(1-\alpha\right)
\end{equation}
gives a radial temperatue profile
\begin{equation}
    T_d = A_H^{1/4}\left[\left(\frac{R_H}{R}\right)^{2}T_H^4 + \left(\frac{R_{ext}}{D}\right)^{2}T_{ext}^4\right]^{1/4}\left(1-\alpha\right)^{1/4}
    \label{equn:simpleRadEq}
\end{equation}
This is shown for a Trappist-1 type star at various distances from a star similar to $\theta^1$C ({$R_{ext}=10\,R_\odot, T_{ext} = 39000\,$K}) in Figure \ref{fig:Tprofiles_simple}. 

For an ensemble of external sources this can be generalized to
\begin{equation}
    T_d = A_H^{1/4}\left[\left(\frac{R_H}{R}\right)^{2}T_H^4 + \sum_k\left(\frac{R_{ext,k}}{D_k}\right)^{2}T_{ext,k}^4\right]^{1/4}\left(1-\alpha\right)^{1/4}
   \label{equn:simpleRadEqAgg}    
\end{equation}
Note that in practice most discs will not be optically thin and the albedo is a function of wavelength and the grain size distribution.

The mass estimate given by equation \ref{equn:massEstimate} in the Rayleigh-Jeans regime scales inversely with the Temperature, so if assuming a temperature of 20\,K the overestimate of the disc mass  is by a factor $T_d/20$. If the external source dominates the bulk of the disc temperature this would hence lead to an overestimate that scales as $D^{-1/2}$.

\section{Radiative Transfer Models}
\label{sec:RT}
To explore the dust temperature of externally irradiated discs in more detail we ran Monte Carlo radiative transfer calculations using the \textsc{torus} code \citep{2019A&C....27...63H}. The approach used is based on that of \cite{1999A&A...344..282L}. The code uses an adapative mesh refinement grid-based approach, with a 2D cylindrical ($R,z$) grid geometry and the host star at the origin. The total source luminosity (host star plus any external source) is discretised into photon packets that undertake a physically motivated random walk over the grid  until they escape. This contributes to the energy density in each cell which sets the dust temperature. 

In these calculations, photon packets from the host star are emitted with random direction and a frequency randomly sampled from the stellar spectrum. When an external photon source is included, it is done so beyond the upper bound of the grid. External photons are hence introduced in a plane parallel fashion from the upper bound of the grid, with frequency again randomly sampled from the external source's spectrum. We choose this approach rather than an isotropic external source, since many irradiated discs close to massive stars are clearly not symmetrically irradiated. This way, we can ask questions like whether the external radiation field affects just the upper atmosphere, also the disc mid plane, or even the far side of the disc from the external radiation source. In this paper we only consider a plane parallel external field propagating parallel to the $z$ axis on the cylindrical $R,z$ grid, since there is a higher probability of the radiation impinging primarily upon the disc surface than the disc edge for a random disc orientation.  Given that 3D models would be required for arbitrary external radiation orientations (otherwise you end up with what is effectively a cylindrical external source) we argue that this situation is the prudent choice.

\subsection{Disc density distribution and dust properties}
\label{sec:discanddust}
Our calculations will solve for the dust radiative equilibrium temperature, but density and grain distributions need to be specified. 

We construct the disc using a truncated power law with a surface density profile
\begin{equation}
    \Sigma(R) = \Sigma_{1\textrm{au}}\left(\frac{R}{\textrm{au}}\right)^{-1}
\end{equation}
and scale height of the form 
\begin{equation}
    H(R) = H_{1\textrm{au}}\left(\frac{R}{\textrm{au}}\right)^{m}
\end{equation}
where we choose $m = 1$. {The inner disc radius is 0.1\,au in each case.} For such a surface density profile the mass encapsulated scales linearly with the radius, and the surface density normalization is given by
\begin{equation}
    \Sigma_{1\textrm{au}} = \frac{M_d}{2\pi R_{d} \times (1\,\textrm{au})}
\end{equation}
{where $R_d$ is the disc outer radius, beyond which the density is set to a negligibly low value ($10^{-35}$\,g\,cm$^{-3}$)}. 
The density structure on the cylindrical grid is then described by
\begin{equation}
    \rho(R,z) = \frac{\Sigma(R)}{\sqrt{2\pi}H(R)}\exp\left(-\frac{z^2}{2H(R)^2}\right). 
\end{equation}
A summary of the parameters of discs for the models used in this paper is given in Table \ref{tab:models}. {We include a mixture of larger discs (100\,au) and smaller ones that are more representative of continuum disc sizes in the ONC \citep[10-50\,au,][]{2018ApJ...860...77E}}. 

For the exploration in this paper we use a single grain population. These are \cite{1984ApJ...285...89D} silicates with a minimum and maximum grain size of 0.1\,$\mu$m and 2\,mm respectively. The power law of the $dn(a)/da \propto a^{-q}$ distribution is $q=3.3$.

\begin{figure*}
    \centering
    \includegraphics[width=\columnwidth]{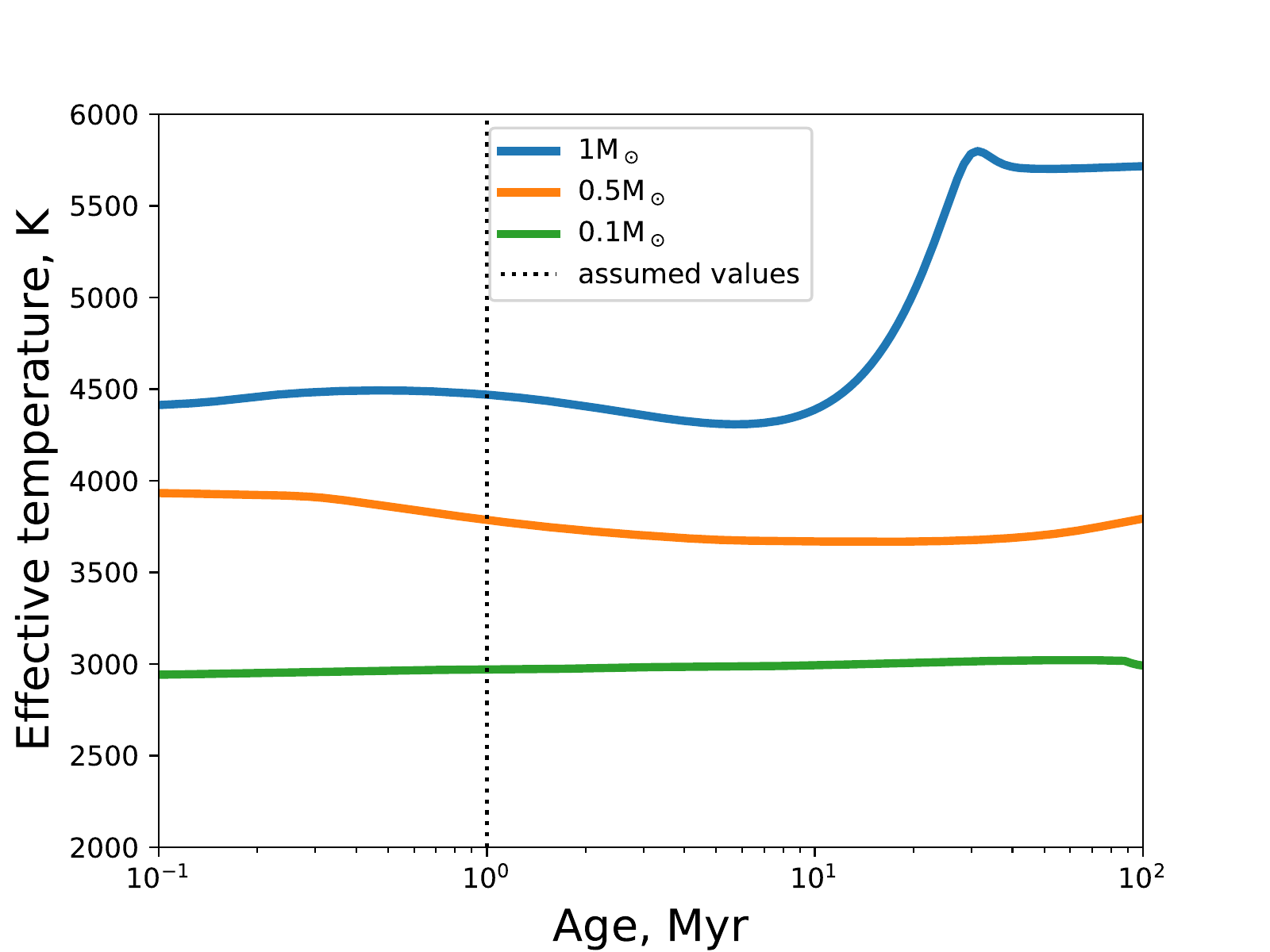}
    \includegraphics[width=\columnwidth]{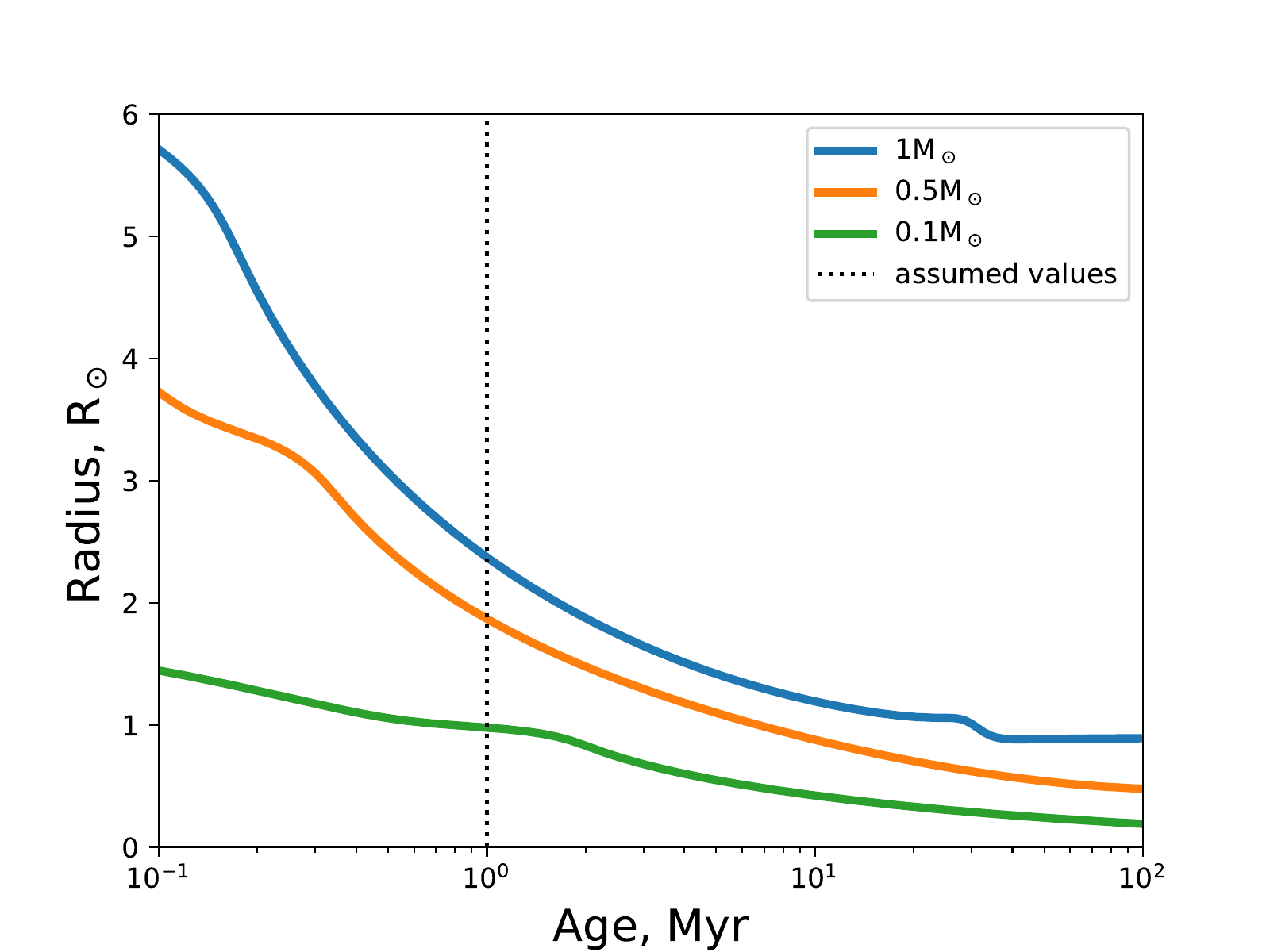}    
    \caption{MIST v1.2 temperature and radius evolutionary tracks for stars of different masses. We use the values at 1\,Myr (the vertical dotted line).  }
    \label{fig:stellarEvo}
\end{figure*}

We do not solve for the hydrostatic equilibrium structure of the disc in these models. Rather, a parametric disc structure is imposed and the radiative equilibrium temperature calculated {this caveat is discussed in section \ref{sec:caveats}}. We consider a few different values for the scale height at 1\,au in the solar mass case, from the canonical $H/R=0.1$ down to $H/R=0.025$ to represent a more settled case. If optically thin the dust radiative equilibrium temperature of an isolated disc scales with the host star luminosity as $T_d\propto L_*^{1/4}$. At lower stellar masses we hence scale the imposed $H/R=0.1$ by a factor
\begin{equation}
    f_H = \left(\frac{L_*}{L_\odot}\right)^{1/8}\left(\frac{M_*}{M_\odot}\right)^{-1/2}.
\end{equation}
We note that we do not attempt to parameterise an external envelope in these models since the bulk of the dust remains within the disc \citep[e.g.][]{2020MNRAS.492.1279S}. Dust in an envelope will have been liberated from the disc in a wind so is just a redistribution of a small fraction of the material in our model, where we are focusing on the disc itself. This does however mean that we cannot model things like infrared emission, to which the proplyd envelope makes an important contribution \citep[e.g.][]{2017A&A...604A..69C}. 

We place each model star-disc at 1, 0.5, 0.1, 0.05 and 0.01\,pc from a $\theta^1$C analogue, with radius 10$\,R_\odot$ and effective temperature of 39000\,K. We also include an isolated version of each model disc with no external source.

\begin{table}
    \centering
    \begin{tabular}{ccccccccccccc}
         \hline
         Stellar  &  $T_*$ & $R_*$ &   $R_d$  & $\Sigma_{1\textrm{au}}$ & Disc dust   & $H_{1\textrm{au}}$  \\
        Mass ($M_\odot$) & (K) & ($R_\odot$) & (au) & g\,cm$^{-2}$ &  mass ($M_\oplus$) & (au) \\
         \hline
        1& 4468 & 2.35 & 100  & 0.425 &10.0 & 0.1  \\
        1& 4468 & 2.35 & 100  & 0.425 &10.0 &0.05  \\
        1& 4468 & 2.35 & 100  & 0.425 &10.0 &0.025 \\
        1& 4468 & 2.35 & 50  & 0.851 &10.0 &0.1\\
        1& 4468 & 2.35 & 50  & 0.851 &10.0 &0.05\\
        1& 4468 & 2.35 & 50  & 0.851 &10.0 &0.01\\
        1& 4468 & 2.35 & 25  & 1.70 &10.0 &0.1\\
        1& 4468 & 2.35 & 10  & 4.25 &10.0 &0.1 \\
        0.5& 3784 & 1.86 & 50 & 0.851 &10.0 & 0.12 \\
        0.5& 3784 & 1.86 & 25 & 1.70 &10.0 & 0.12\\
        0.5& 3784 & 1.86 & 10 & 4.25 &10.0 & 0.12\\
        0.5& 3784 & 1.86 & 10 & 2.12 &5.0 & 0.12\\
        0.5&  3784 & 1.86 & 10 & 0.425 &1.0 & 0.12\\
        0.1&  2970 & 0.98 & 50 & 0.851 &10.0 & 0.21 \\
        0.1&  2970 & 0.98 & 25 & 1.70 &10.0 & 0.21 \\
        0.1&  2970 & 0.98 & 10 &4.25 &10.0 & 0.21 \\
        0.1& 2970 & 0.98 & 10 &2.12 &5.0 & 0.21 \\
        0.1&  2970 & 0.98 & 10 & 0.425 &1.0 & 0.21 \\
        \hline
    \end{tabular}
    \caption{Summary of the star-disc properties of models in this paper. Each of these models are placed at different distances from an external UV source.  }
    \label{tab:models}
\end{table}

\subsection{Defining the sources}
\label{sec:sources}
The stellar luminosity provides an additional complication because the pre-main sequence luminosity is a different function of time for different stellar masses and needs to be computed with a model such as those by \cite{2000A&A...358..593S} or MESA \citep{2011ApJS..192....3P, 2013ApJS..208....4P, 2015ApJS..220...15P, 2016ApJ...823..102C, 2016ApJS..222....8D}. In particular, before the zero age main sequence the luminosity is a shallower function of stellar mass. For this initial study we adopt stellar temperature and radii taken from {Mesa Isochrones and Stellar Tracks (MIST, v1.2) web interpolator} \footnote{\url{http://waps.cfa.harvard.edu/MIST/interp_tracks.html}} at a time of 1Myr, assuming Fe/H = 0 {(the default, using a protosolar Fe/H=0.0142 changes both the stellar temperature and radius by less than 0.2\,per cent)} and initial v/v$_\textrm{crit}$ = 0.4. The adopted stellar temperatures and radii are shown in Figure \ref{fig:stellarEvo} in the wider context of the stellar evolution. Table \ref{tab:models} also specifies the stellar parameters used. 

An argument could be made about the short lifetimes of discs in close proximity to massive stars (the proplyd lifetime problem) but this problem is coupled to uncertainties on the disc properties, such as mass, age and time actually spent in the high UV environment. We therefore deem this choice of PMS stellar properties at a single age to be pragmatic and a much better representation of the probable reality than zero age main sequence luminosities. 

There is a further consideration required, which is how to treat the external radiation field entering the 2D cylindrical grid. In section \ref{sec:RT} we argued for treating this as a plane parallel field entering from the upper bound of the model grid. However, on a cylindrical grid the cylindrical area of a cell centred on $R_i$ with thickness $\Delta R$ increases as 
\begin{equation}
    \Delta A = 2 \pi \left[ \left(R_i+\frac{\Delta R}{2}\right)^2 - \left(R_i-\frac{\Delta R}{2}\right)^2\right]
\end{equation}
so randomly introducing photon packets at the upper grid boundary with uniform probability in $R$ would not correspond to a uniform planar field, but rather one with a density per unit area that decreases with $R$. To address this, the square root of the randomly sampled $R/R_{grid}$ is used, where $R_{grid}$ is the outermost radial coordinate of the grid. 

The energy carried per photon packet $\Delta E$ is always the same, with different frequencies simply corresponding to different numbers of photons in the packet. For $N$ packets and total source luminosity $L$ the energy (per unit time) carried by a packet is $\delta E = L/N$. For the purpose of computing the photon packet energy, the luminosity of the external source, which is intrinsically $L_{ext,0}$, is scaled by 
\begin{equation}
    L_{ext} = L_{ext,0}\frac{\pi R_{grid}^2}{4\pi D_{grid}^2}
\end{equation}
where $R_{grid}$ is the radial grid size and $D_{grid}$ is the distance of the external source from the grid. 

\begin{figure*}
    \centering
    \includegraphics[width=\columnwidth]{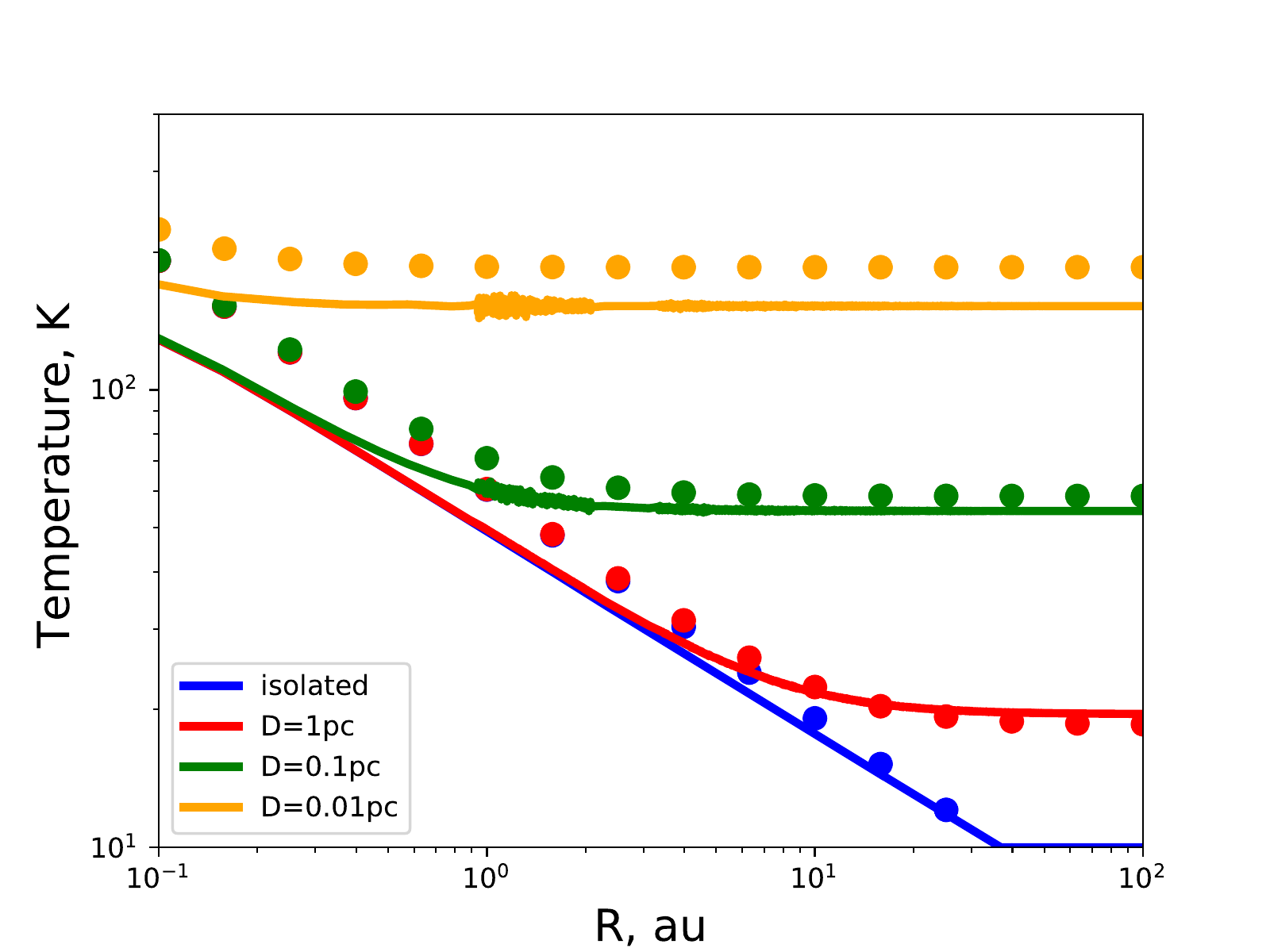}      
    \includegraphics[width=\columnwidth]{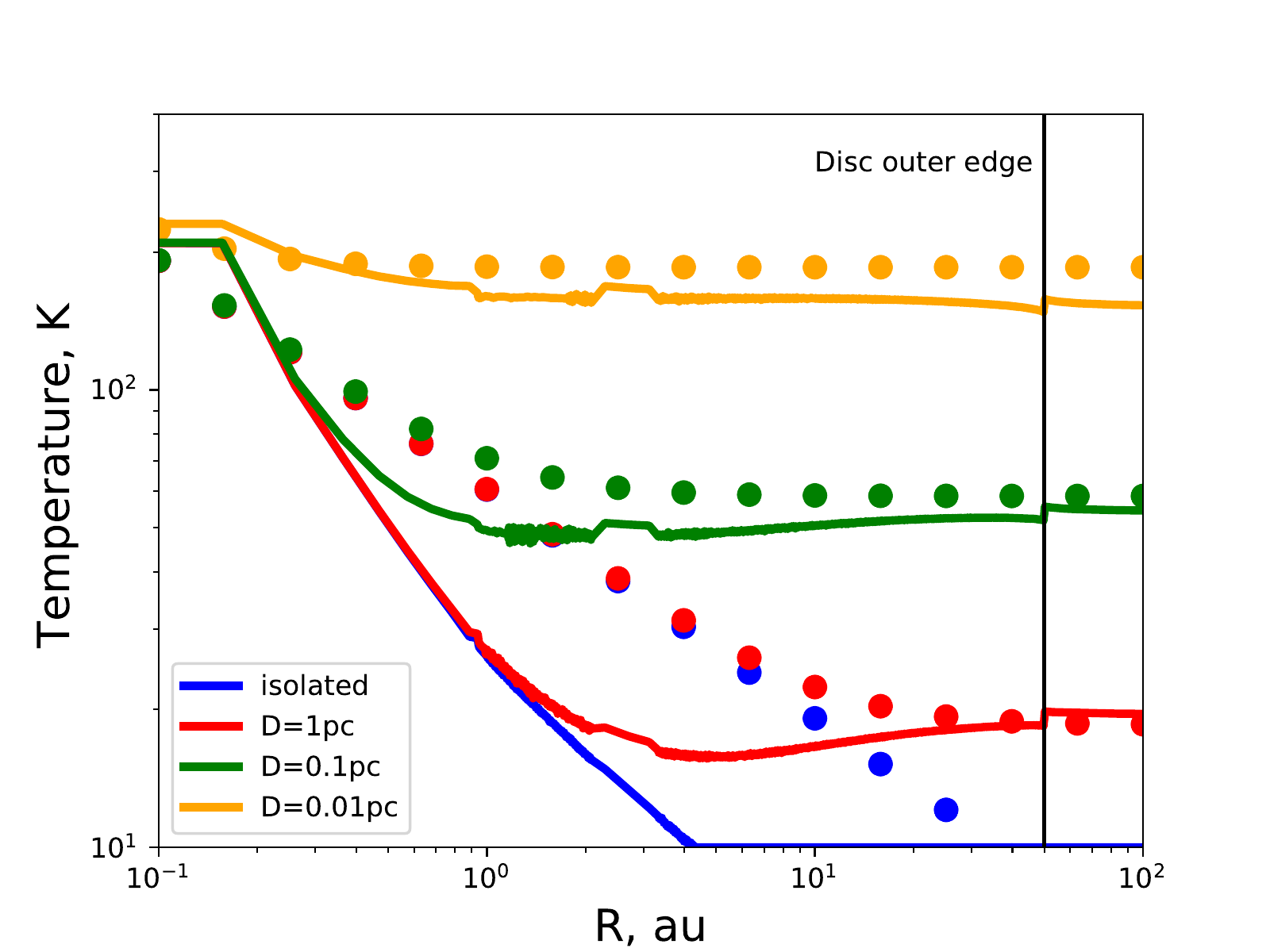}
    \caption{Points represent analytic {optically thin} disc radiative equilibrium {mid-plane} radial temperature profiles for a host star with $T_H=2550$, $R_H=0.121$ (Trappist-1) at various distances from a star similar to $\theta^1$C. The solid lines are results from \textsc{torus} radiative transfer models. The left hand panel is an extremely optically thin disc ($M_d=10^{-17}\,$M$_\oplus$) and the right hand panel is for a $R_d=$50\,au, $M_d=10\,$M$_\oplus$ disc. }
    \label{fig:Tprofiles}
\end{figure*}

\subsection{Obvious caveats}
\label{sec:caveats}
Before proceeding to look at calculation results we note some of the obvious caveats with our approach to the modelling. This is not designed to be a detailed study of grain/disc evolution and planet formation, but rather a first look at the effect of external irradiation on the thermal structure of the circumstellar {millimetre dust} disc. To this end we consider only a single snapshot of any given disc. We consider a distribution of grain sizes that is the same everywhere (i.e. there is no midplane larger grain population and diffuse atmospheric population due to settling and/or entrainment of smaller grains in a wind). We also do not solve for hydrostatic equilibrium as {these are essentially models of the dust, and the combination of growth/settling/drift with a photoevaporative wind that entrains small grains means that a} hydrostatic model is not necessarily  better than a parametric dust structure. However, we do explore the impact of the the scale height of the dust which can be considered a rough proxy for settling. For the systems in close proximity to a strong external radiation field we are assuming that the embedded dust is still geometrically disc-like, despite there being a larger scale proplyd structure. For this reason we do not explore synthetic SEDs in this paper, since the warm envelope makes an important contribution to the SED \citep[e.g.][]{2002ApJ...578..897R, 2017A&A...604A..69C}.
We will address the dynamic evolution of the dust in future work. 

Another caveat is that although we consider a realistic pre main sequence stellar luminosity, we are not including the effects of accretion luminosity which may occasionally increase the heating from the central source.

\subsection{Benchmarking in the optically thin regime}
We begin by comparing \textsc{torus} against equation \ref{equn:simpleRadEq} in the extremely optically thin regime. We still use a disc model, as described above, only with a dust mass of $10^{-17}$\,M$_\oplus$. In this first look we use host star properties similar to those of Trappist-1, i.e. $T_H=2550\,$K and $R_H=0.121$\,R$_\odot$ (though note that in our later calculations we use appropriate pre main sequence stellar properties from evolutionary models as described above) and the external source is similar to $\theta^1$C ($T_{ext}$=39000\,K, $R_{ext}=10\,$R$_\odot$). 

A comparison of equation \ref{equn:simpleRadEq} and the mid-plane temperature of this extremely optically thin disc is shown in the left hand panel of Figure \ref{fig:Tprofiles}. The right hand panel is the same setup, only with the disc mass increased to a more typical value of $10$\,M$_\oplus$. For the analytic approximation in this case we assume a negligible albedo. 

Overall the optically thin simulations give good agreement with the analytic solution, particularly beyond 10\,au. When a more realistic disc mass is used the agreement is still good in the outer disc (dependent upon the distance of the disc from the radiation source). As expected, in the more realistic disc mass case the optically thick inner disc drops to a lower temperature value than equation \ref{equn:simpleRadEq} predicts. 

We also study the deviation from the simple analytic solution in our other models in section \ref{sec:thermalStructure}, finding that the analytic approximation is typically good in the regions of the disc where the external field sets the temperature, but typically overestimates the disc temperature by a factor of around 4 where the host star dominates. Nevertheless equation \ref{equn:simpleRadEq} could still be used to place upper limits on the temperature and therefore corresponding lower limits on the disc mass.

\subsection{Testing against ONC proplyd disc temperature estimates}

\cite{2013MNRAS.430.3406T} estimated the disc dust temperature of three proplyds in the ONC: LVC 2, HS1 and HST10. They also provided disc mass/radius estimates and the projected separation from $\theta^1$\,C. We used their disc/separation parameters, employing the host star parameters and canonical scale height at 1\,au that we use for a $0.5$\,M$_\odot$ star (see table \ref{tab:models}) to run comparison models of these systems. 

\begin{figure}
    \centering
    \includegraphics[width=\columnwidth]{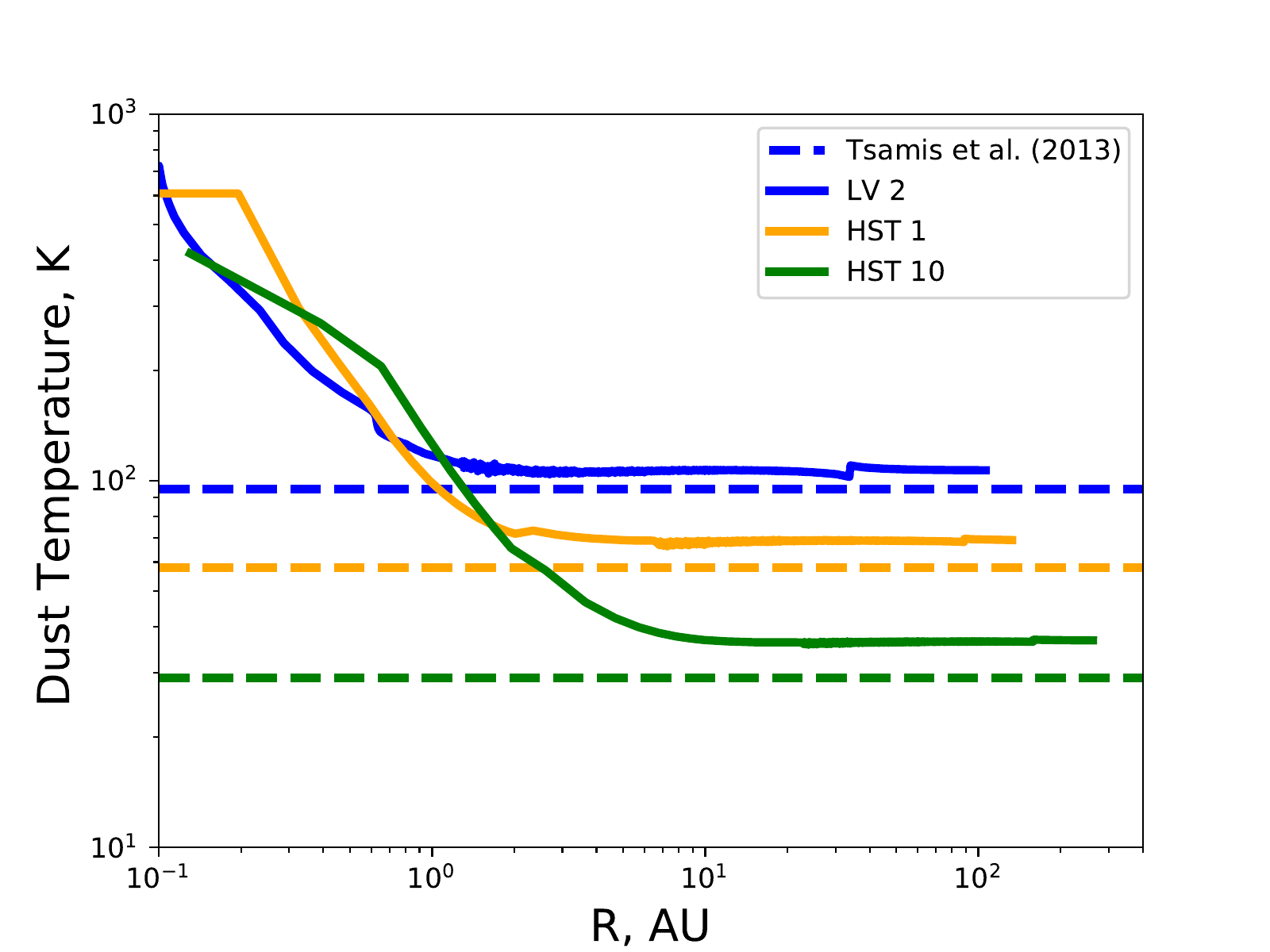}
    \caption{Comparison of our mid-plane dust temperature structure with the disc dust temperatures inferred by \protect\cite{2013MNRAS.430.3406T} for the ONC proplyds LV2, HST 1 and HST 10. The agreement in the outer disc where the external radiation dominates is to within 20\,per cent. }
    \label{fig:Tsamis}
\end{figure}

The \cite{2013MNRAS.430.3406T} dust temperature estimates were made by assuming that 25\,per cent of the $\theta^1$\,C bolometric luminosity reaches the disc surface. A comparison of their inferred dust temperatures with our model mid-plane dust temperature structure is given in Figure \ref{fig:Tsamis}. In the outer disc where the external field dominates the temperature the agreement is to within at worst 20\,per cent.

\section{Results and discussion}
We begin by presenting the thermal structure of externally irradiated dust discs, before turning our attention to observables and effect upon disc dust mass estimates. 

\subsection{Thermal structure of internally and externally irradiated discs}
\label{sec:thermalStructure}
\begin{figure*}
    \includegraphics[width=2.1\columnwidth]{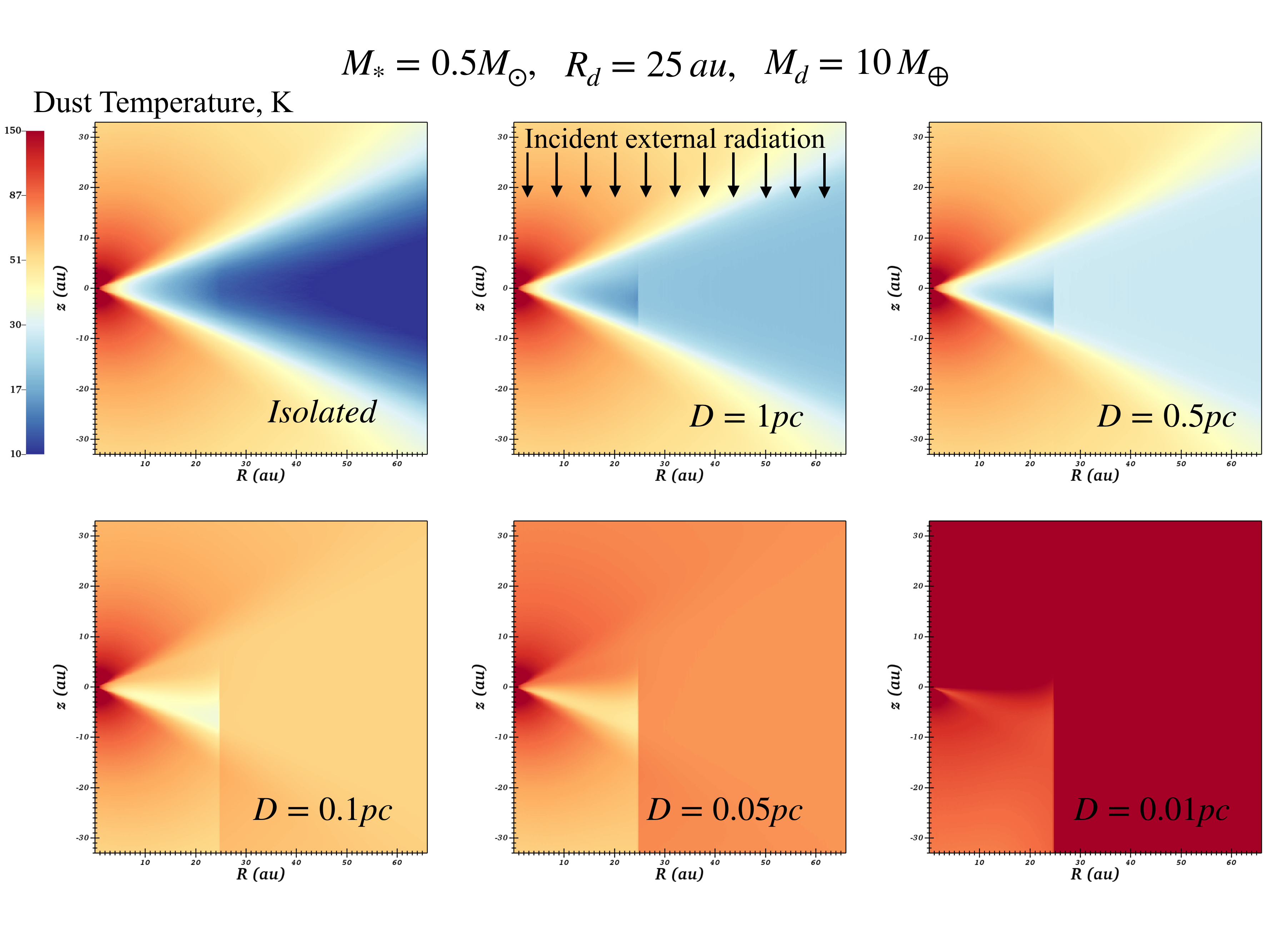}
    \vspace{-1cm}
    \caption{The dust radiative equilibrium temperature structure of a 10\,M$_\oplus$ dust mass, 25\,au disc around a 0.5\,M$_\odot$ star at various distances from a 39000\,K 10\,R$_\odot$ star. The top left panel is isolated (no external source) and in the other panels the label in the lower right denotes the distance from the external source which irradiates the grid from the top part of each panel, as illustrated in the top-middle panel.  }
    \label{fig:MainPlot}
\end{figure*}

Figure \ref{fig:MainPlot} shows the dust temperature structure of one of our model discs at different distances from a $\theta^1$C-like external source. This scenario is a 0.5\,M$_\odot$ host star with a 25\,au, 10\,M$_\oplus$ dust disc and the result is representative of the typical behaviour of our models. The upper left panel is an isolated disc, so there is no external radiation source, and illustrates the motivation for the usual assumption of a disc temperature of 20\,K when computing disc mass estimates. The other panels from left to right, top to bottom, move the disc closer to an external radiation field, which is impinging from the upper boundary. This results in the upper half of the disc being warmer than the lower side, though the lower side is still heated relative to the isolated disc case. Since we only consider a static snapshot we do not assess the dynamical impact of this asymmetric heating here. 

\begin{figure}
    \centering
    \includegraphics[width=\columnwidth]{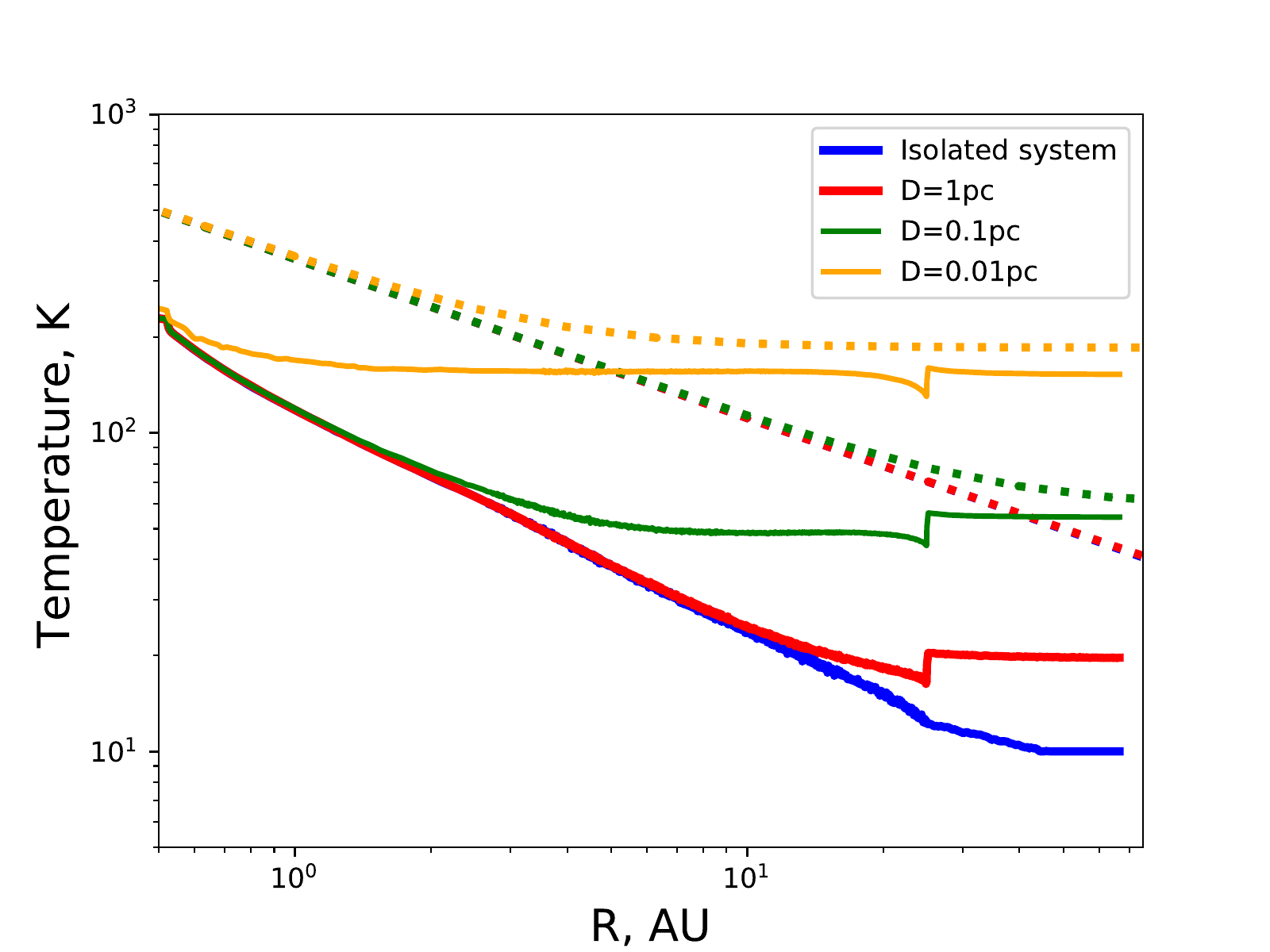}
    \vspace{-0.3cm}
    \includegraphics[width=\columnwidth]{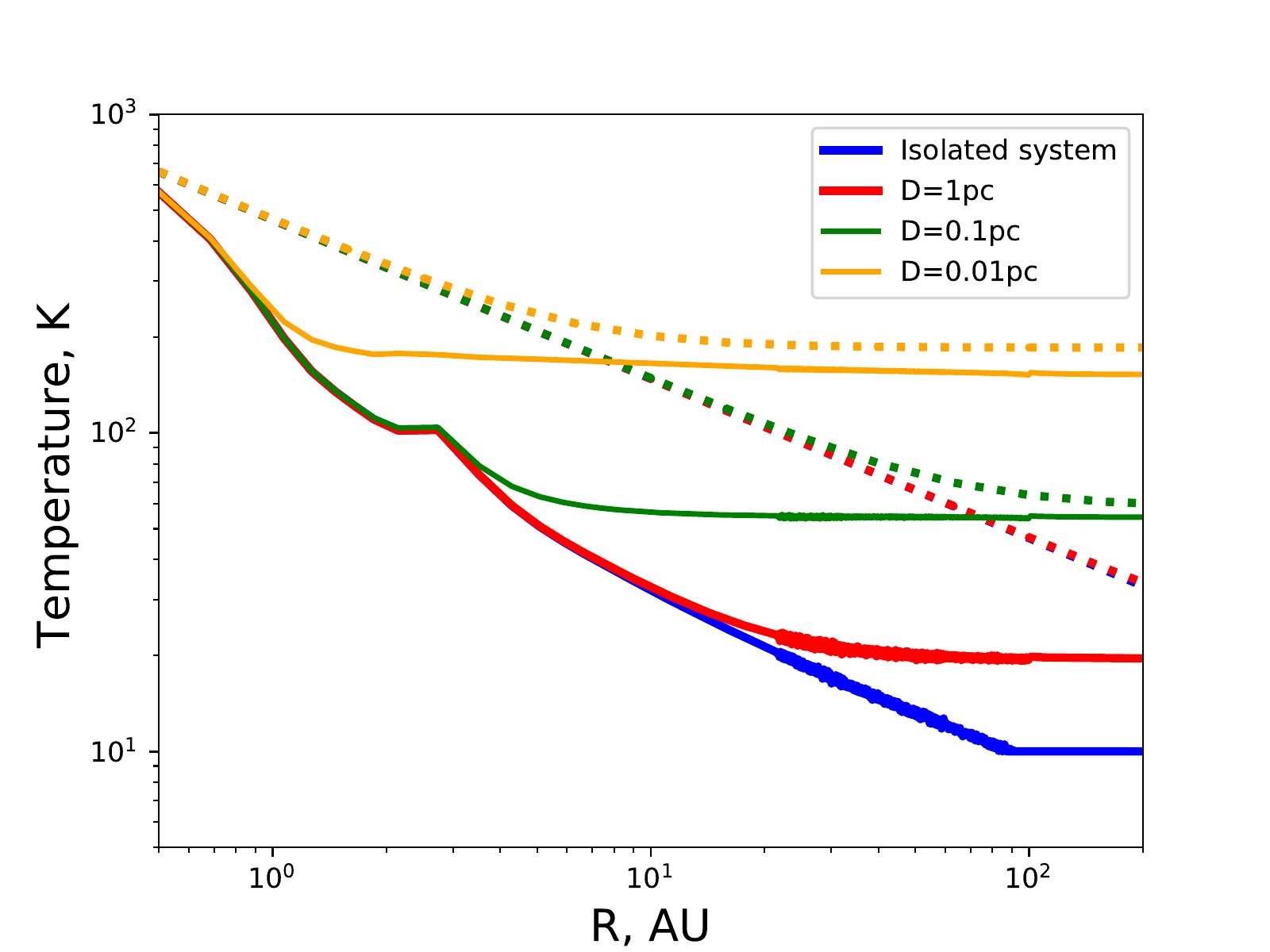}    
    \caption{Mid-plane temperature profiles of externally irradiated disc models. The upper panel has a 0.5\,M$_\odot$ host star, 25\,au disc outer radius and 10\,M$_\oplus$ of dust. The lower panel has a 1\,M$_\odot$ host star, 100\,au disc outer radius and 10\,M$_\oplus$ of dust. The solid lines are the radiative transfer calculation results and the dotted lines are the simple analytic radiative equilibrium approximations. }
    \label{fig:midPlaneTemperatures}
\end{figure}

Figure \ref{fig:midPlaneTemperatures} shows the mid-plane temperature of two of our star-disc models. In each case we include the isolated result, as well as at distances of 1, 0.1 and 0.01\,pc from the external source. The models are {M$_*=0.5\,$M$_\odot$}, R$_d=25$\,au, $M_d=10\,$M$_\oplus$ (upper panel) and {M$_*=1\,$M$_\odot$}, R$_d=100$\,au, $M_d=10\,$M$_\oplus$ (lower panel), and are representative of the typical behaviour we see in all of the models. In addition to the model results (solid lines) we also include the analytic approximation given by equation \ref{equn:simpleRadEq} (dashed lines). For the discs nearest the external source (0.01\,pc) the analytic approximation is reasonable but there is more significant deviation (a factor $\sim$4) in the inner parts of the disc at larger separations when the external source doesn't set the disc temperature. 

A key point to make here is that the disc temperature is in excess of 20\,K once a disc is at a separation of less than a parsec from the $\theta^1$ C like source. We will assess how this affects disc mass estimates in section \ref{sec:DustMasses}, but note now that an upper limit on the temperature (which could be provided by the analytic approximation) corresponds to a lower limit on the inferred mass using equation \ref{equn:massEstimate}, which is a valuable addition to an estimate made assuming a temperature of 20\,K.

\subsubsection{Dependence of temperature structure on dust scale height}
The main calculations in this paper assume a single mixed grain population with small (0.1\,$\mu$m) grains right the way through to 2\,mm sized grains. In reality the larger grains settle towards the mid-plane and some of the smaller grains may be lifted upwards in a photoevaporative wind. To make a simple assessment of the possible impact of these processes we ran calculations with smaller scale heights,  but otherwise identical parameters (stellar properties, disc mass/radius). 

\begin{figure}
    \centering
    \includegraphics[width=\columnwidth]{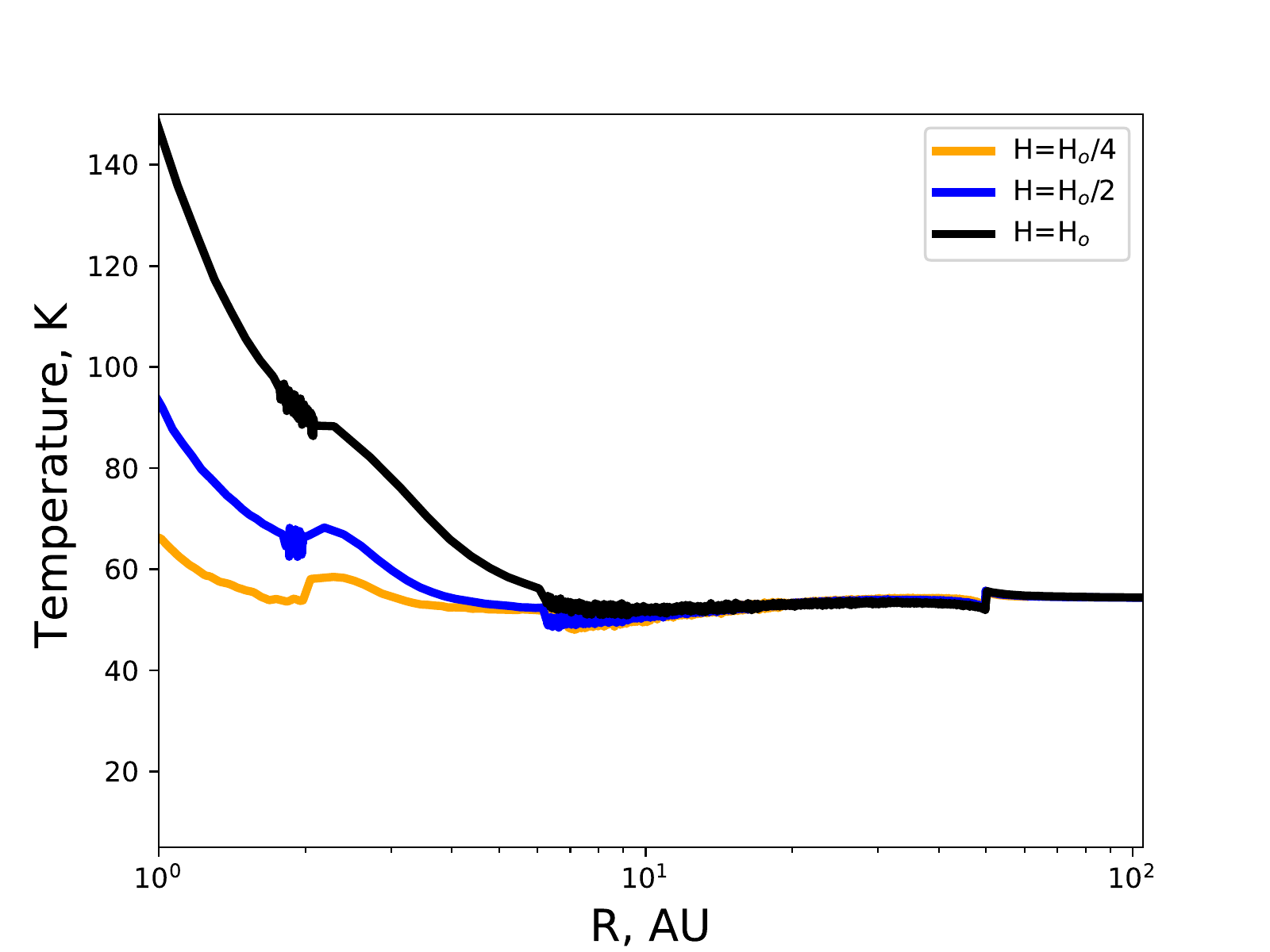}
    \caption{Mid-plane temperature profiles of externally irradiated disc models with different scale heights at a distance of 0.1\,pc from the external source. A smaller scale height results in lower inner-disc temperatures, with the external radiation field dominating the mid-plane temperature at smaller distances from the host star.   This is because the mid-plane is more optically thick to the host-star radiation, but the overall vertical column is unchanged.  }
    \label{fig:midTScaleHeight}
\end{figure}

A comparison of the mid-plane temperature structure of a canonical scale height model and others with a factor two and four smaller scale heights is given in Figure \ref{fig:midTScaleHeight}. The temperature of the outer parts of the disc heated by the external source doesn't change, but the radius at which the external source dominates (in the mid-plane at least) moves inwards. This is because the mid-plane is denser and so becomes optically thick to the host star radiation more quickly. However, the vertical column at any given radius (and hence to the external radiation) is the same. 

There are two important points to make from this comparison. The first is that although we are considering a single density distribution for each model, there is only a finite amount of dust in the model so even if small grains are elevated away from the mid-plane we do not expect it to significantly affect the importance of external heating on the disc dust temperature (e.g. if the system were a proplyd). The second point is that what lifting dust away from the mid-plane may do is increase the ability of the host star to contribute to warming the disc (and vice-versa for settling). 

\begin{figure}
    \centering
    \includegraphics[width=\columnwidth]{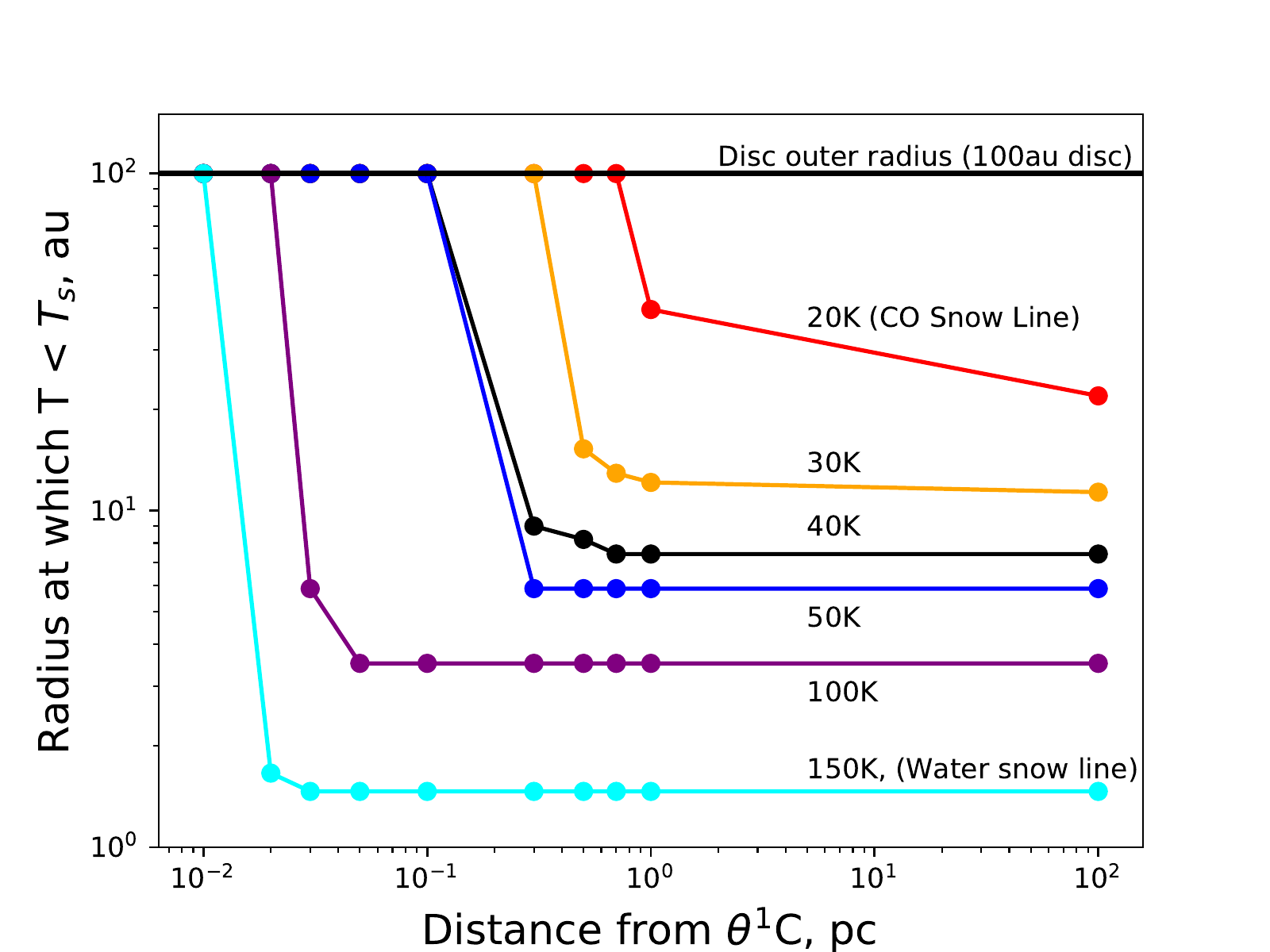}
    \caption{The radius at which certain temperatures are reached in the disc mid-plane. For example the CO snow line is at 22\,au until about 1\,pc distance from a $\theta^1$C analogue, but at closer distances no CO snow line would exist. The water snow line, of possible importance to inner planet formation, is relatively unaffected by environment except for at very close separations. This is for a 100\,au disc with a $10\,M_\oplus$ dust mass and $H/R=0.1$ }
    \label{fig:Tsnowlines}
\end{figure}

\subsubsection{Comments on the effect of external irradiation upon snow lines}
Another important effect of external irradiation is the impact upon the locations (or even existence of) snow lines \citep[{e.g. as illustrated in a single value of the external UV field in}][]{2013ApJ...766L..23W}. The idea that planetary composition could link to formation locations in the disc through snow lines \citep{2011ApJ...743L..16O} has been complicated in recent years due to factors like time dependent chemistry and pebble drift \citep{2019MNRAS.487.3998B}, thermal instability at snow lines \citep{2020MNRAS.495.3160O} and the time evolution of the host star luminosity \citep{2021MNRAS.500.4658M}. External irradiation adds yet another factor. 

Here we do not study the time evolution of snow lines like e.g. \cite{2019MNRAS.487.3998B} and \cite{2021MNRAS.500.4658M}, but in Figure \ref{fig:Tsnowlines} we show the radius in a 100\,au, 10\,M$_\oplus$ disc around a 1\,M$_\odot$ star at which the temperature drops below certain values as a function of distance from the external radiation source. The rightmost set of points represent the thermal structure of an isolated disc (placed at 100\,pc in Figure \ref{fig:Tsnowlines}). Then moving left from this rightmost point one can follow how the location of particular mid-plane temperatures migrates outwards. 

A temperature of 20\,K roughly corresponds to the temperature of the CO snow line, 50\,K is roughly the temperature of the CO$_2$ snow line and 150\,K is roughly the temperature of the water snow line. The CO snow line moves outwards significantly compared to an isolated disc, doing so by a factor two even at separations of about a parsec from the O star. At closer separations, the CO snow line ceases to exist \citep[][{inferred no CO snow line for their single externally irradiated disc at what would have been $D<0.1\,$pc}]{{2013ApJ...766L..23W}}. The water snow line has been proposed as an important part of the planet forming mechanism for ensembles of planets on small orbits such as in Trappist-1 \citep{2017A&A...604A...1O, 2019A&A...627A.149S}. Naively based on our results we would anticipate that this would be rather resilient to the effect of the external irradiation as the water snow line is only significantly affected at separations of $\leq10^{-2}$\,pc. Of course even if the water snow line location isn't affected, the mass reservoir for planet formation in Trappist-1 like systems could still be significantly affected by external photoevaporation \citep{2018MNRAS.475.5460H}.

\subsection{Dust mass estimates}
\label{sec:DustMasses}
As dicussed above, {trends in disc dust masses near massive stars are assessed using equation \ref{equn:massEstimate}, restated here}
\begin{equation*}
    M_d = \frac{D^2F_\nu}{\kappa_\nu B_\nu(T)}
\end{equation*}
where $D$ is the source distance, $F_\nu$ the measured flux, $\kappa_\nu$ the opacity and $B_\nu(T)$ the Planck function. In the Rayleigh-Jeans regime there is a linear scaling of the Planck function with the temperature \citep[][assumed $T=20$\,K]{2017AJ....153..240A, 2018ApJ...860...77E}.

The significant additional heating from external sources considered here could therefore conceivably lead to overestimated disc dust masses {which might suppress any imprint of external photoevaporation on the disc mass}. To test this we produce 850\,$\mu$m synthetic observations from our calculations, again using the Monte Carlo radiative transfer approach in \textsc{torus}. {We assume a distance of 414\,pc in our synthetic observations.} {In the Monte Carlo radiative transfer calculation the opacity is as described in section \ref{sec:discanddust}}. {We also have to assume an opacity to compute the dust mass. We follow \cite{2018ApJ...860...77E} and use {of $\kappa_\nu = 3.1$\,cm$^2$\,g$^{-1}$ at 850\,$\mu$m} for this. }

{It is important to note that the opacity is also a source of significant uncertainty when estimating disc masses. Millimetre opacity estimates can vary by an order of magnitude, particularly depending on whether grain growth is accounted for, which corresponds to around an order of magnitude uncertainty on the disc mass \citep{1993Icar..106...20M, 2004A&A...416..179N,2010A&A...512A..15R}}. {Based on the discussion that follows below, in the ONC an order of magnitude overestimate of the dust mass due to external heating is only achieved at around 0.01\,pc from $\theta^1 C$, so the uncertainty in disc mass due to the opacity is generally going to be larger than that due to external heating.  On the other hand, here we are mainly concerned with the effect of heating on estimated disc masses as a function of projected separation from a strong radiation source, i.e. the key thing is the variation in the opacity between sources, not uncertainty in its actual value. In such a scenario, where the discs are in the same region and similarly aged, it is reasonable to expect that the degree of grain growth will be fairly similar and the millimetre opacity will not be changing by an order of magnitude on a disc-to-disc basis.} 

 We do not account for any instrumentation effects such as finite beam size or interferometric effects as we are mostly interested in the impact of assuming a single temperature when computing the mass regardless of distance from the external source, which is insensitive to such factors. 

\begin{figure}
    \centering
    \includegraphics[width=1.1\columnwidth]{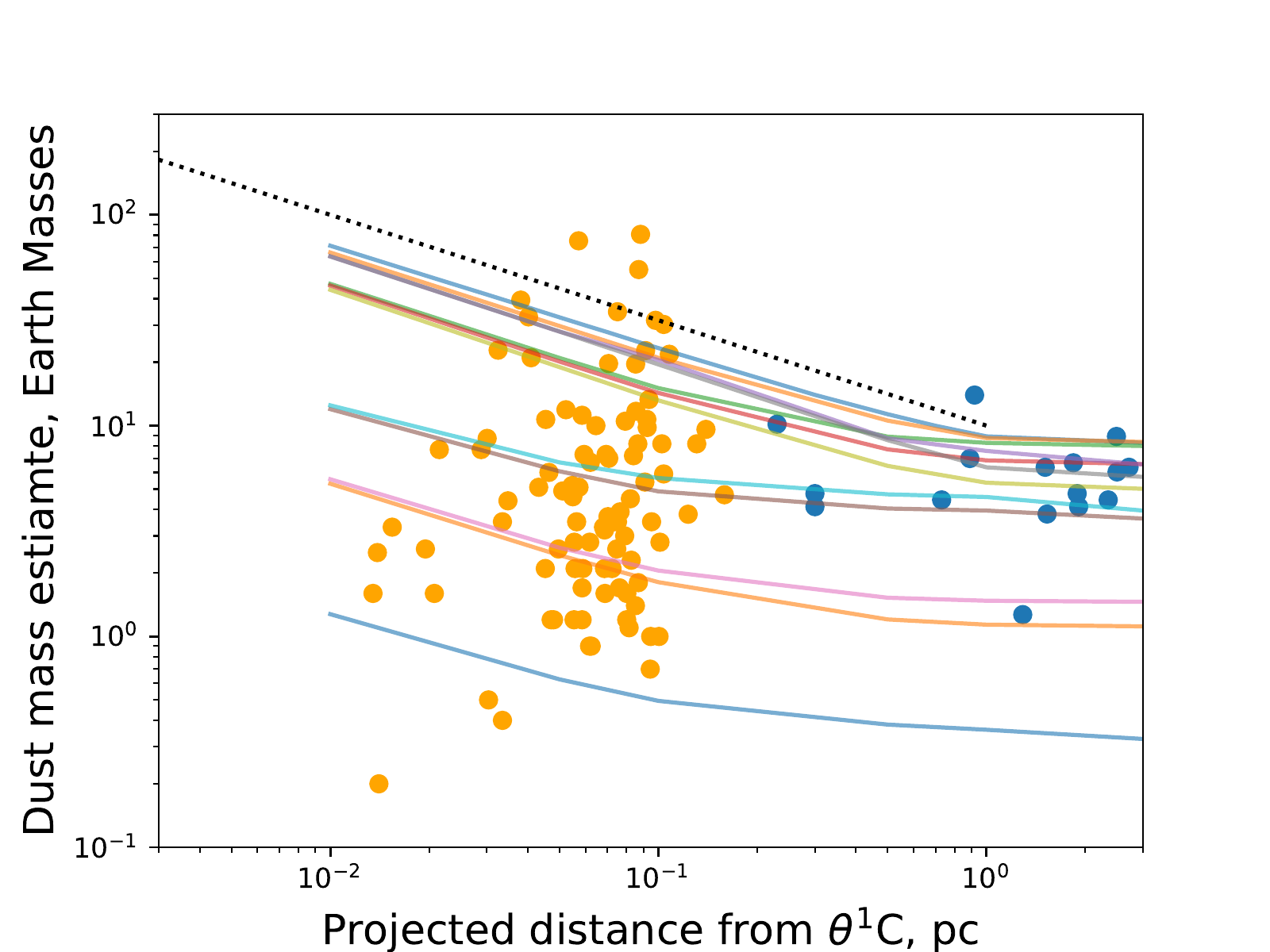}
    \caption{Inferred disc 850\,$\mu$m dust masses as a function of projected separation from $\theta^1 C$. The blue points are detections (we ignore upper limits here) from \protect\cite{2014ApJ...784...82M} and the orange points are detections from \protect\cite{2018ApJ...860...77E}. The solid coloured lines are \textsc{torus} radiative equilibrium synthetic 850\,$\mu$m models, with tracks representing an identical star-disc, just at different distances from the external radiation source. As the star-disc approaches the massive external source the mass of the dust disc is overestimated when assuming a temperature of 20\,K, since the true dust temperature is in fact higher. The dotted line denotes a $D^{-1/2}$ scaling. }
    \label{fig:theta1C}
\end{figure}

Figure \ref{fig:theta1C} shows the estimated dust masses as a function of projected separation from $\theta^1$ C from \cite{2014ApJ...784...82M} and \cite{2018ApJ...860...77E}. Overplotted are a series of lines corresponding to the masses inferred from our synthetic observations, showing how the inferred disc mass changes for any given star-disc system as it is moved closer to the external source. As expected the inferred disc mass increases nearer to the external source due to the increased temperature, with an overestimate of the disc mass by a factor $\sim 10$ at a separation of 0.01\,pc. We also find that the overestimate scales with distance from the external source as  $D^{-1/2}$ as expected from the discussion in section \ref{sec:simpleRadEq}, and significant deviations from the 20\,K assumption start at about 1\,pc (in this particular case). The $D^{-1/2}$ scaling interior to 1\,pc is illustrated by the dotted line in Figure \ref{fig:theta1C}. 

\begin{figure}
    \centering
    \includegraphics[width=\columnwidth]{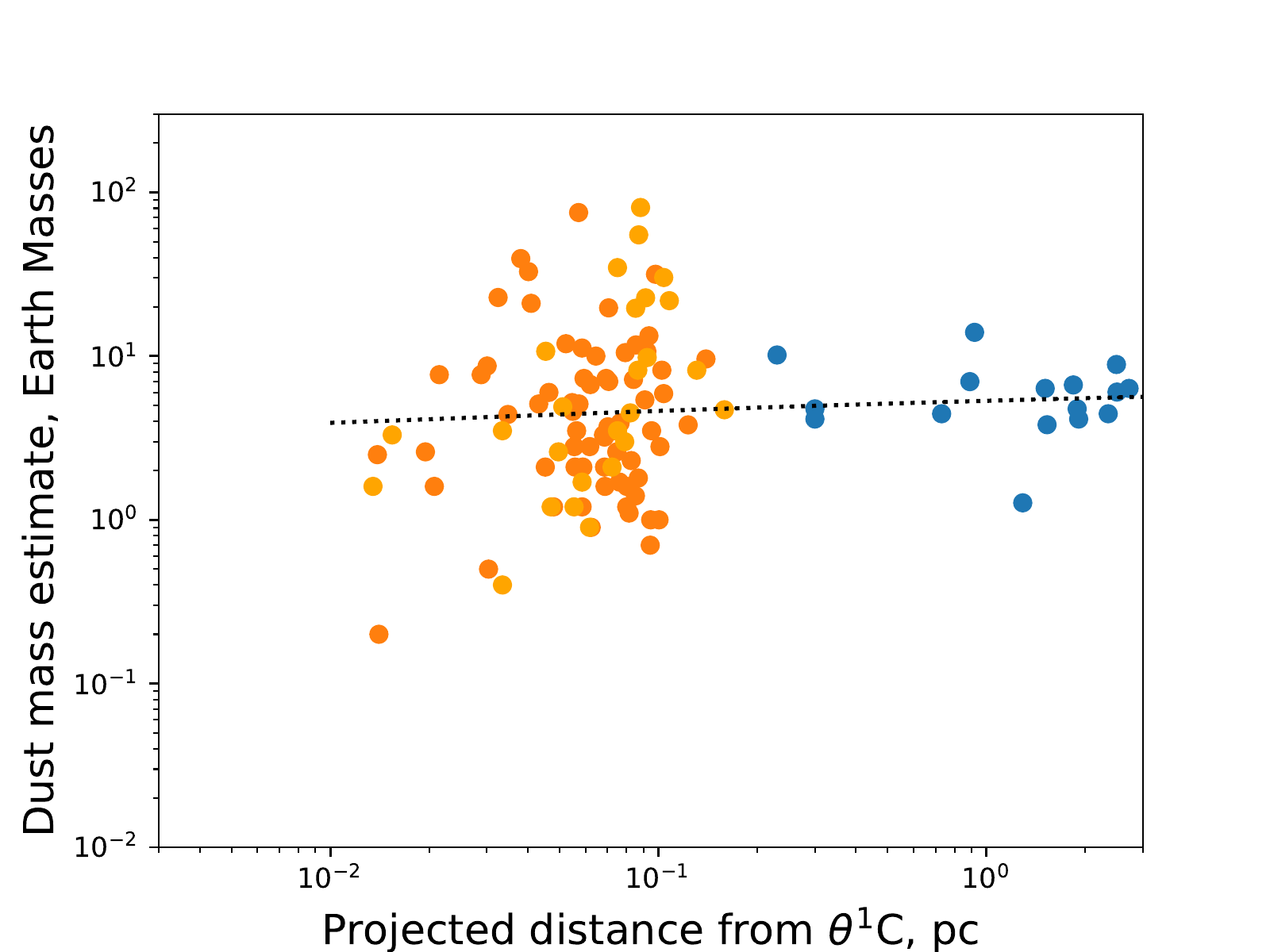}
    \includegraphics[width=\columnwidth]{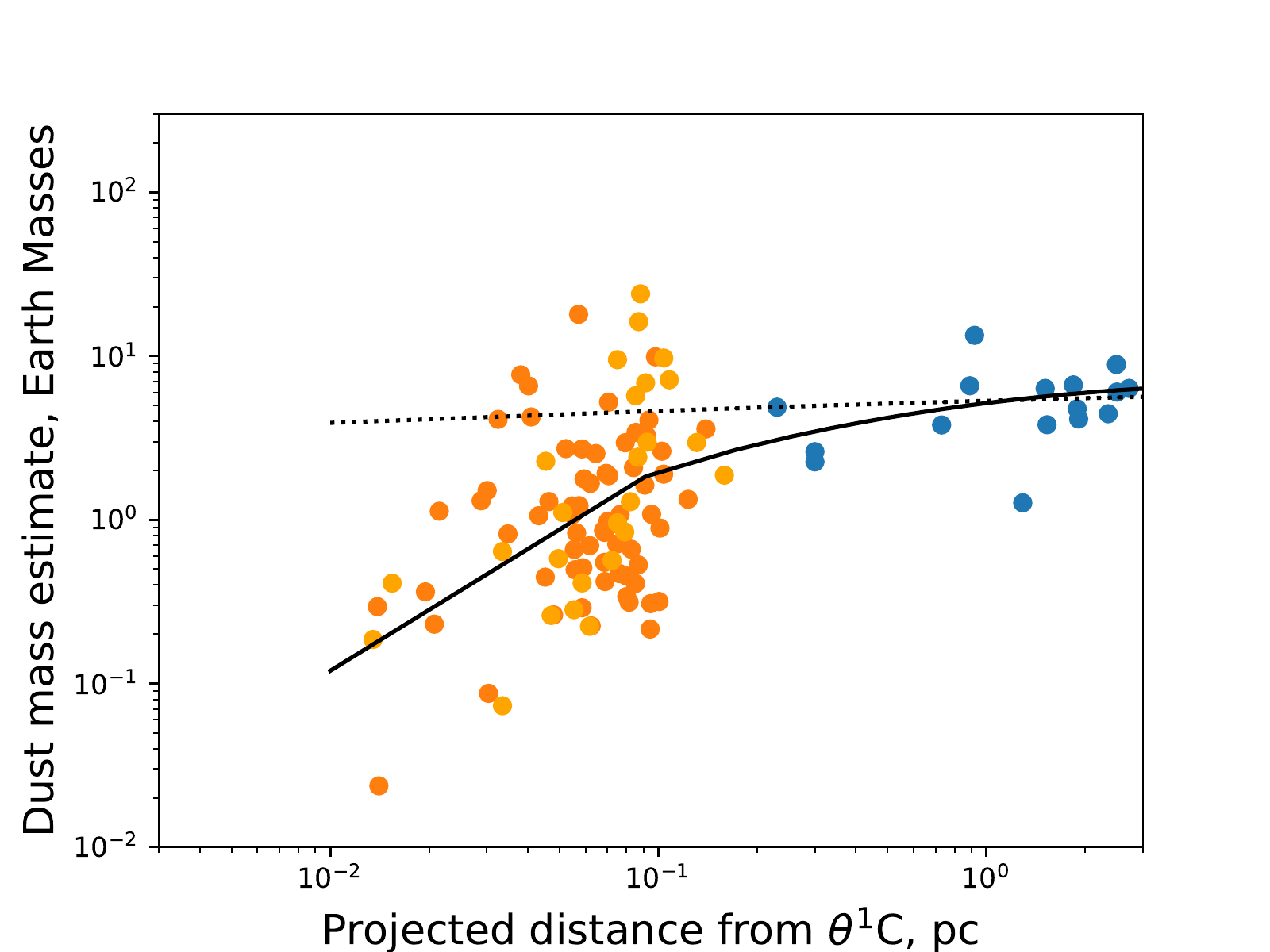}    
    \caption{Inferred disc dust masses as a function of projected separation from $\theta 1 C$. The upper panel is a combination of the detections (we ignore upper limits here) from \protect\cite{2014ApJ...784...82M} (blue points) and \protect\cite{2018ApJ...860...77E} (orange points). The lower panel is scaled by the $(D/\textrm{pc})^{0.5}$ factor by which our models suggests the masses are being overestimated. {The dotted and solid lines approximate the moving average of the upper and lower distributions respectively}.    }
    \label{fig:theta1CScaled}
\end{figure}

Ideally each individual source in the \cite{2014ApJ...784...82M} and \cite{2018ApJ...860...77E} surveys would be subject to bespoke modelling, but for a first assessment of the possible impact of this bias we can simply scale the observed masses by the overestimate factor inferred from our models $f = \left(D/\textrm{pc}\right)^{-1/2}$, which we do in Figure \ref{fig:theta1CScaled}. The upper panel shows the original data and the lower panel accounting for the temperature-induced overestimate of the disc mass by multiplying by the factor $f$. {The dotted and solid lines approximate the moving average in each original and scaled data respectively}. Of course there is uncertainty as to the true separation (since all we plot in Figure \ref{fig:theta1CScaled} is the projected separation), but our results suggest that a trend in disc mass with separation may be being suppressed. 

It is worth noting that a trend in disc dust masses as a function of projected separation is not necessarily expected in an external photoevaporation scenario. The dynamical evolution of a star cluster on such small spatial scales is relatively fast, so any such signature might not survive very long. On the other hand it may also be quickly imprinted. 

\subsubsection{Effect on other surveys and observations}
Here we have focused our attention on the ONC and $\theta ^1$\,C. Another notable example of disc dust masses as a function of projected separation is in the older $\sigma$ Orionis cluster by \cite{2017AJ....153..240A}, where an increase in disc dust mass as a function of projected separation from $\sigma$ Ori was observed {(with a stronger correlation than observed in the ONC)}.
{The distance $D$ from a single source at which a disc would transition to being externally heated at a radius $R$ in the disc is}
\begin{equation}
    D = R\left(\frac{L_{ext}}{L_{H}}\right)^{1/2}
    \label{equn:D}
\end{equation}
{where $L_{ext}$ and $L_H$ are the external and host star luminosities.}

The correlation of disc mass with separation in $\sigma$ Orionis was at separations of more like beyond a parsec from $\sigma$ Ori and the luminosity of $\sigma$ Ori is also about a factor 5 weaker than $\theta ^1$\,C. We hence expect any overestimate of disc masses as a function of separation due to external heating plays a much weaker role in that region than in the ONC. \cite{2020A&A...640A..27V} also surveyed disc masses with ALMA towards NGC 2024 {at projected separations from the O8V star IRS 2b comparable to the separations of discs considered from $\theta ^1$\,C in the ONC (down to 0.01\,pc) by \cite{2018ApJ...860...77E}}. {IRS 2b is also around a factor 5 less luminous than $\theta ^1$\,C, however the square root dependency on this in equation \ref{equn:D} means that external heating should be important for disc heating/mass estimates out to separations of around 0.5\,pc in that region.  }

{Outside of regions like those discussed above within around a parsec of massive stars we do not expect that it is important to consider external heating of the dust disc for mass estimates, especially given the order of magnitude uncertainty on the opacity \citep[see section \ref{sec:DustMasses} and][]{1993Icar..106...20M, 2004A&A...416..179N,2010A&A...512A..15R}. } {However it is also worth noting that the bolometric luminosity has a shallower decline with stellar mass than the FUV, so depending on the cluster density the aggregate of lower mass  close neighbours may also lead to substantial heating beyond the reaches of massive stars. Stellar luminosities and 3D positions using gaia distances \citep[e.g.][]{2016A&A...595A...1G, 2016A&A...595A...2G, 2018A&A...616A...1G} could be combined with equation \ref{equn:simpleRadEqAgg} (or equation \ref{equn:D}) to make a first assessment of this. }

\subsubsection{Effect of disc inclination and position of the external radiation source }
Estimating disc dust masses using equation \ref{equn:massEstimate} assumes that the emission is optically thin, in which case there should not be a sensitivity to inclination. If the disc were optically thick then the situation becomes substantially more complicated. On the one hand the disc mass may be being underestimated if it is optically thick. The disc mass inferred would then also be sensitive to the disc inclination, since if the warmer irradiated side were presented to the observer the emission would be brighter and (if a temperature of 20\,K were assumed) some of the mass estimate deficit from being optically thick accounted for. Conversely if the cold side were presented the inferred mass would be lower. For simplicity it is best to assume that the disc is generally optically thin and consider the inclination/optical depth effects when undertaking bespoke modelling of individual discs.

\subsection{Possible effect on grain evolution and planet formation}
So far we have focused on the effect that the increased disc dust temperature has on continuum mass estimates and commented on the effect on snow lines. It is also worth briefly considering the possible impact of higher disc dust temperatures and a flatter radial temperature profile on grain evolution. 

Firstly and most simply, our results indirectly support the work of \cite{2018MNRAS.474..886N}, where higher outer disc temperatures were imposed in planet population synthesis models. They found that this outer heating is important for suppressing populations of cold Jupiters, particularly at low metallicity, which are not observed. 

We can also consider the impact of externally heated disc temperature structures on grain growth and drift following the simple model of \cite{2012A&A...539A.148B}. For example the drift timescale and maximum grain size before the onset of radial drift both scale inversely with the square of the sound speed and log pressure gradient ($\left|\textrm{d}\ln(P)/ \textrm{d}\ln(R)\right|$). For a heated irradiated disc both of these quantities will be decreased by a factor $T_{\textrm{isolated}}/T_{\textrm{irradiated}}$ due to the change in sound speed. Since the temperature profile in the externally heated disc is much flatter, the reduced log pressure gradient will also act to slightly further reduce the drift timescale and maximum grain size for drift. So overall drift may happen to smaller grains, and more rapidly. Whether this helps to promote or further hinder planet formation (in addition to the photoevaporative depletion of the disc) remains to be explored. 

Another factor is that the Stokes number (grain size) for fragmentation of grains through collisions scales inversely with the temperature, so the Stokes number/size for fragmentation will be decreased by a factor $T_{\textrm{isolated}}/T_{\textrm{irradiated}}$ in a warmed irradiated disc. Once again, this could act to further suppress planet formation in tandem with the removal of material by external photoevaporation. 

Finally is worth noting that if the disc temperature distribution is flatter (as it is in the bulk of an irradiated disc) then it is easier to introduce pressure bumps through density perturbations which could lead to dust trapping. 

{It is important to note here that the total fraction planet hosts for which the dust discs would have been exposed to external heating at some point in their lifetime is uncertain and cannot be assessed by counting the instantaneous fraction of stars in nearby regions within $\sim$1\,pc of O stars. Stellar clusters are dynamically evolving systems, with ongoing star formation and stars moving in and out of high radiation parts of a cluster \citep[e.g.][]{2019MNRAS.490.5478W}. In addition, most stars formed at around a redshift of 2 \citep{2014ARA&A..52..415M} when star forming regions may typically have resembled more extreme/massive clusters with strong radiation fields \citep[e.g.][]{2020SSRv..216...69A}.  }

\section{Summary and conclusions}
We use Monte Carlo radiative transfer models to study the radiative equilibrium structure of protoplanetary discs irradiated by both the host star and an external source. {In particular we are interested in how external irradiation affects trends in disc mass estimates as a function of separation from massive stars compared to  assuming a constant temperature (as is often the case in recent ALMA continuum surveys) since these are used to infer the impact of external photoevaporation on disc evolution.} We draw the following main conclusions from this work. \\

\noindent 1) The majority of the dust component of discs in the vicinity ($<1\,$pc) of massive stars can be heated by the external radiation field to well in excess of the 20\,K {recently assumed in  ALMA millimetre continuum surveys} for estimating disc masses {in these high radiation environments}. This leads to disc masses being overestimated when assuming a 20\,K disc, since warmer discs are brighter. The overestimate scales with the separation from the external source as roughly $D^{-1/2}$, until the disc is sufficiently far from the external source that the host star  dominates the temperature structure. In the vicinity of $\theta^1$C in the ONC the external field starts to play an important role at a separation of about a parsec and results in roughly a factor 10 overestimate on the dust mass at separations of 0.01\,pc. \\

\noindent 2) Applying a simple $D^{-1/2}$ scaling to observed disc dust masses as a function of projected separation from $\theta^1$C (at separations less than 1\,pc, as motivated by our models) results in a significantly stronger variation in disc mass with projected separation. Though of course this has the caveat that projected separation is not necessarily the true separation. Bespoke {SED modelling} of ONC discs on a case-by-case basis would be required for a more rigorous assessment of the impact of external heating on disc mass estimates. \\

\noindent 3) External heating {from nearby massive stars can} affect mid-plane snow line locations, which further complicates the idea of being able to relate planet compositions to formation radii. The CO snow line is quite readily removed completely from discs near massive stars, whereas the water snow line (which may be important for inner planet formation) is resilient to cluster heating except for very small separations ($\leq0.01$\,pc) from a $\theta^1$C type external source. {The fraction of stars subject to this kind of heating is is yet to be determined.}  \\

\noindent 4) {Generally when estimating the mass of any given disc the uncertainty on the opacity will dominate over that due to external heating. The main importance of external heating is when searching for trends in disc properties in the vicinity of massive stars and this paper provides the tools for making a first assessment of this}. 

\section{Data availability}
The \textsc{torus} code is publicly available\footnote{\url{https://bitbucket.org/tjharries/torus/}}. The observational data included is available directly from \cite{2014ApJ...784...82M} and \cite{2018ApJ...860...77E}. All plotting scripts and any other information are available on request. 

\section*{Acknowledgements}
We thank the referee for their careful assessment of this paper. We also thank Andrew Sellek, Andrew Winter, {Josh Eisner}, {Nick Ballering} and Will Henney for useful comments and discussions. TJH is funded by a Royal Society Dorothy Hodgkin Fellowship. This work used the DiRAC@Durham facility managed by the Institute for Computational Cosmology on behalf of the STFC DiRAC HPC Facility (www.dirac.ac.uk). The equipment was funded by BEIS capital funding via STFC capital grants ST/P002293/1, ST/R002371/1 and ST/S002502/1, Durham University and STFC operations grant ST/R000832/1. DiRAC is part of the National e-Infrastructure.

\appendix

\bsp

\bibliographystyle{mnras}
\bibliography{molecular.bib} 

\begin{thebibliography}{}
\makeatletter
\relax
\def\mn@urlcharsother{\let\do\@makeother \do\$\do\&\do\#\do\^\do\_\do\%\do\~}
\def\mn@doi{\begingroup\mn@urlcharsother \@ifnextchar [ {\mn@doi@}
  {\mn@doi@[]}}
\def\mn@doi@[#1]#2{\def\@tempa{#1}\ifx\@tempa\@empty \href
  {http://dx.doi.org/#2} {doi:#2}\else \href {http://dx.doi.org/#2} {#1}\fi
  \endgroup}
\def\mn@eprint#1#2{\mn@eprint@#1:#2::\@nil}
\def\mn@eprint@arXiv#1{\href {http://arxiv.org/abs/#1} {{\tt arXiv:#1}}}
\def\mn@eprint@dblp#1{\href {http://dblp.uni-trier.de/rec/bibtex/#1.xml}
  {dblp:#1}}
\def\mn@eprint@#1:#2:#3:#4\@nil{\def\@tempa {#1}\def\@tempb {#2}\def\@tempc
  {#3}\ifx \@tempc \@empty \let \@tempc \@tempb \let \@tempb \@tempa \fi \ifx
  \@tempb \@empty \def\@tempb {arXiv}\fi \@ifundefined
  {mn@eprint@\@tempb}{\@tempb:\@tempc}{\expandafter \expandafter \csname
  mn@eprint@\@tempb\endcsname \expandafter{\@tempc}}}

\bibitem[\protect\citeauthoryear{{Adamo} et~al.,}{{Adamo}
  et~al.}{2020}]{2020SSRv..216...69A}
{Adamo} A.,  et~al., 2020, \mn@doi [\ssr] {10.1007/s11214-020-00690-x}, \href
  {https://ui.adsabs.harvard.edu/abs/2020SSRv..216...69A} {216, 69}

\bibitem[\protect\citeauthoryear{{Andre} \& {Montmerle}}{{Andre} \&
  {Montmerle}}{1994}]{1994ApJ...420..837A}
{Andre} P.,  {Montmerle} T.,  1994, \mn@doi [\apj] {10.1086/173608}, \href
  {https://ui.adsabs.harvard.edu/abs/1994ApJ...420..837A} {420, 837}

\bibitem[\protect\citeauthoryear{{Andrews} \& {Williams}}{{Andrews} \&
  {Williams}}{2005}]{2005ApJ...631.1134A}
{Andrews} S.~M.,  {Williams} J.~P.,  2005, \mn@doi [\apj] {10.1086/432712},
  \href {https://ui.adsabs.harvard.edu/abs/2005ApJ...631.1134A} {631, 1134}

\bibitem[\protect\citeauthoryear{{Andrews} \& {Williams}}{{Andrews} \&
  {Williams}}{2007}]{2007ApJ...671.1800A}
{Andrews} S.~M.,  {Williams} J.~P.,  2007, \mn@doi [\apj] {10.1086/522885},
  \href {https://ui.adsabs.harvard.edu/abs/2007ApJ...671.1800A} {671, 1800}

\bibitem[\protect\citeauthoryear{{Ansdell}, {Williams}, {Manara}, {Miotello},
  {Facchini}, {van der Marel}, {Testi}  \& {van Dishoeck}}{{Ansdell}
  et~al.}{2017}]{2017AJ....153..240A}
{Ansdell} M.,  {Williams} J.~P.,  {Manara} C.~F.,  {Miotello} A.,  {Facchini}
  S.,  {van der Marel} N.,  {Testi} L.,   {van Dishoeck} E.~F.,  2017, \mn@doi
  [\aj] {10.3847/1538-3881/aa69c0}, \href
  {https://ui.adsabs.harvard.edu/abs/2017AJ....153..240A} {153, 240}

\bibitem[\protect\citeauthoryear{{Ansdell} et~al.,}{{Ansdell}
  et~al.}{2020}]{2020AJ....160..248A}
{Ansdell} M.,  et~al., 2020, \mn@doi [\aj] {10.3847/1538-3881/abb9af}, \href
  {https://ui.adsabs.harvard.edu/abs/2020AJ....160..248A} {160, 248}

\bibitem[\protect\citeauthoryear{{Beckwith}, {Sargent}, {Chini}  \&
  {Guesten}}{{Beckwith} et~al.}{1990}]{1990AJ.....99..924B}
{Beckwith} S. V.~W.,  {Sargent} A.~I.,  {Chini} R.~S.,   {Guesten} R.,  1990,
  \mn@doi [\aj] {10.1086/115385}, \href
  {https://ui.adsabs.harvard.edu/abs/1990AJ.....99..924B} {99, 924}

\bibitem[\protect\citeauthoryear{{Biganzoli}, {Potenza}  \&
  {Robberto}}{{Biganzoli} et~al.}{2017}]{2017ApJ...840...55B}
{Biganzoli} D.,  {Potenza} M. A.~C.,   {Robberto} M.,  2017, \mn@doi [\apj]
  {10.3847/1538-4357/aa6bf9}, \href
  {https://ui.adsabs.harvard.edu/abs/2017ApJ...840...55B} {840, 55}

\bibitem[\protect\citeauthoryear{{Birnstiel}, {Klahr}  \&
  {Ercolano}}{{Birnstiel} et~al.}{2012}]{2012A&A...539A.148B}
{Birnstiel} T.,  {Klahr} H.,   {Ercolano} B.,  2012, \mn@doi [\aap]
  {10.1051/0004-6361/201118136}, \href
  {https://ui.adsabs.harvard.edu/abs/2012A&A...539A.148B} {539, A148}

\bibitem[\protect\citeauthoryear{{Booth} \& {Ilee}}{{Booth} \&
  {Ilee}}{2019}]{2019MNRAS.487.3998B}
{Booth} R.~A.,  {Ilee} J.~D.,  2019, \mn@doi [\mnras] {10.1093/mnras/stz1488},
  \href {https://ui.adsabs.harvard.edu/abs/2019MNRAS.487.3998B} {487, 3998}

\bibitem[\protect\citeauthoryear{{Boyden} \& {Eisner}}{{Boyden} \&
  {Eisner}}{2020}]{2020ApJ...894...74B}
{Boyden} R.~D.,  {Eisner} J.~A.,  2020, \mn@doi [\apj]
  {10.3847/1538-4357/ab86b7}, \href
  {https://ui.adsabs.harvard.edu/abs/2020ApJ...894...74B} {894, 74}

\bibitem[\protect\citeauthoryear{{Champion}, {Bern{\'e}}, {Vicente}, {Kamp},
  {Le Petit}, {Gusdorf}, {Joblin}  \& {Goicoechea}}{{Champion}
  et~al.}{2017}]{2017A&A...604A..69C}
{Champion} J.,  {Bern{\'e}} O.,  {Vicente} S.,  {Kamp} I.,  {Le Petit} F.,
  {Gusdorf} A.,  {Joblin} C.,   {Goicoechea} J.~R.,  2017, \mn@doi [\aap]
  {10.1051/0004-6361/201629404}, \href
  {https://ui.adsabs.harvard.edu/abs/2017A&A...604A..69C} {604, A69}

\bibitem[\protect\citeauthoryear{{Choi}, {Dotter}, {Conroy}, {Cantiello},
  {Paxton}  \& {Johnson}}{{Choi} et~al.}{2016}]{2016ApJ...823..102C}
{Choi} J.,  {Dotter} A.,  {Conroy} C.,  {Cantiello} M.,  {Paxton} B.,
  {Johnson} B.~D.,  2016, \mn@doi [\apj] {10.3847/0004-637X/823/2/102}, \href
  {http://adsabs.harvard.edu/abs/2016ApJ...823..102C} {823, 102}

\bibitem[\protect\citeauthoryear{{Cleeves}}{{Cleeves}}{2016}]{2016ApJ...816L..21C}
{Cleeves} L.~I.,  2016, \mn@doi [\apjl] {10.3847/2041-8205/816/2/L21}, \href
  {https://ui.adsabs.harvard.edu/abs/2016ApJ...816L..21C} {816, L21}

\bibitem[\protect\citeauthoryear{{Cleeves}, {Adams}  \& {Bergin}}{{Cleeves}
  et~al.}{2013}]{2013ApJ...772....5C}
{Cleeves} L.~I.,  {Adams} F.~C.,   {Bergin} E.~A.,  2013, \mn@doi [\apj]
  {10.1088/0004-637X/772/1/5}, \href
  {https://ui.adsabs.harvard.edu/abs/2013ApJ...772....5C} {772, 5}

\bibitem[\protect\citeauthoryear{{Concha-Ram{\'\i}rez}, {Wilhelm}, {Portegies
  Zwart}  \& {Haworth}}{{Concha-Ram{\'\i}rez}
  et~al.}{2019}]{2019MNRAS.490.5678C}
{Concha-Ram{\'\i}rez} F.,  {Wilhelm} M. J.~C.,  {Portegies Zwart} S.,
  {Haworth} T.~J.,  2019, \mn@doi [\mnras] {10.1093/mnras/stz2973}, \href
  {https://ui.adsabs.harvard.edu/abs/2019MNRAS.490.5678C} {490, 5678}

\bibitem[\protect\citeauthoryear{{Dotter}}{{Dotter}}{2016}]{2016ApJS..222....8D}
{Dotter} A.,  2016, \mn@doi [\apjs] {10.3847/0067-0049/222/1/8}, \href
  {http://adsabs.harvard.edu/abs/2016ApJS..222....8D} {222, 8}

\bibitem[\protect\citeauthoryear{{Draine} \& {Lee}}{{Draine} \&
  {Lee}}{1984}]{1984ApJ...285...89D}
{Draine} B.~T.,  {Lee} H.~M.,  1984, \mn@doi [\apj] {10.1086/162480}, \href
  {http://adsabs.harvard.edu/abs/1984ApJ...285...89D} {285, 89}

\bibitem[\protect\citeauthoryear{{Eisner} et~al.,}{{Eisner}
  et~al.}{2018}]{2018ApJ...860...77E}
{Eisner} J.~A.,  et~al., 2018, \mn@doi [\apj] {10.3847/1538-4357/aac3e2}, \href
  {http://adsabs.harvard.edu/abs/2018ApJ...860...77E} {860, 77}

\bibitem[\protect\citeauthoryear{{Facchini}, {Clarke}  \& {Bisbas}}{{Facchini}
  et~al.}{2016}]{2016MNRAS.457.3593F}
{Facchini} S.,  {Clarke} C.~J.,   {Bisbas} T.~G.,  2016, \mn@doi [\mnras]
  {10.1093/mnras/stw240}, \href
  {http://adsabs.harvard.edu/abs/2016MNRAS.457.3593F} {457, 3593}

\bibitem[\protect\citeauthoryear{{Gaia Collaboration} et~al.,}{{Gaia
  Collaboration} et~al.}{2016a}]{2016A&A...595A...1G}
{Gaia Collaboration} et~al., 2016a, \mn@doi [\aap]
  {10.1051/0004-6361/201629272}, \href
  {https://ui.adsabs.harvard.edu/abs/2016A&A...595A...1G} {595, A1}

\bibitem[\protect\citeauthoryear{{Gaia Collaboration} et~al.,}{{Gaia
  Collaboration} et~al.}{2016b}]{2016A&A...595A...2G}
{Gaia Collaboration} et~al., 2016b, \mn@doi [\aap]
  {10.1051/0004-6361/201629512}, \href
  {https://ui.adsabs.harvard.edu/abs/2016A&A...595A...2G} {595, A2}

\bibitem[\protect\citeauthoryear{{Gaia Collaboration} et~al.,}{{Gaia
  Collaboration} et~al.}{2018}]{2018A&A...616A...1G}
{Gaia Collaboration} et~al., 2018, \mn@doi [\aap]
  {10.1051/0004-6361/201833051}, \href
  {https://ui.adsabs.harvard.edu/abs/2018A&A...616A...1G} {616, A1}

\bibitem[\protect\citeauthoryear{{Garc{\'\i}a-Arredondo}, {Henney}  \&
  {Arthur}}{{Garc{\'\i}a-Arredondo} et~al.}{2001}]{2001ApJ...561..830G}
{Garc{\'\i}a-Arredondo} F.,  {Henney} W.~J.,   {Arthur} S.~J.,  2001, \mn@doi
  [\apj] {10.1086/323367}, \href
  {https://ui.adsabs.harvard.edu/abs/2001ApJ...561..830G} {561, 830}

\bibitem[\protect\citeauthoryear{{Harries}, {Haworth}, {Acreman}, {Ali}  \&
  {Douglas}}{{Harries} et~al.}{2019}]{2019A&C....27...63H}
{Harries} T.~J.,  {Haworth} T.~J.,  {Acreman} D.,  {Ali} A.,   {Douglas} T.,
  2019, \mn@doi [Astronomy and Computing] {10.1016/j.ascom.2019.03.002}, \href
  {https://ui.adsabs.harvard.edu/abs/2019A&C....27...63H} {27, 63}

\bibitem[\protect\citeauthoryear{{Haworth} \& {Clarke}}{{Haworth} \&
  {Clarke}}{2019}]{2019MNRAS.485.3895H}
{Haworth} T.~J.,  {Clarke} C.~J.,  2019, \mn@doi [\mnras]
  {10.1093/mnras/stz706}, \href
  {https://ui.adsabs.harvard.edu/abs/2019MNRAS.485.3895H} {485, 3895}

\bibitem[\protect\citeauthoryear{{Haworth}, {Facchini}, {Clarke}  \&
  {Mohanty}}{{Haworth} et~al.}{2018a}]{2018MNRAS.475.5460H}
{Haworth} T.~J.,  {Facchini} S.,  {Clarke} C.~J.,   {Mohanty} S.,  2018a,
  \mn@doi [\mnras] {10.1093/mnras/sty168}, \href
  {https://ui.adsabs.harvard.edu/abs/2018MNRAS.475.5460H} {475, 5460}

\bibitem[\protect\citeauthoryear{{Haworth}, {Clarke}, {Rahman}, {Winter}  \&
  {Facchini}}{{Haworth} et~al.}{2018b}]{2018MNRAS.481..452H}
{Haworth} T.~J.,  {Clarke} C.~J.,  {Rahman} W.,  {Winter} A.~J.,   {Facchini}
  S.,  2018b, \mn@doi [\mnras] {10.1093/mnras/sty2323}, \href
  {https://ui.adsabs.harvard.edu/abs/2018MNRAS.481..452H} {481, 452}

\bibitem[\protect\citeauthoryear{{Haworth}, {Kim}, {Winter}, {Hines}, {Clarke},
  {Sellek}, {Ballabio}  \& {Stapelfeldt}}{{Haworth}
  et~al.}{2021}]{2021MNRAS.501.3502H}
{Haworth} T.~J.,  {Kim} J.~S.,  {Winter} A.~J.,  {Hines} D.~C.,  {Clarke}
  C.~J.,  {Sellek} A.~D.,  {Ballabio} G.,   {Stapelfeldt} K.~R.,  2021, \mn@doi
  [\mnras] {10.1093/mnras/staa3918}, \href
  {https://ui.adsabs.harvard.edu/abs/2021MNRAS.501.3502H} {501, 3502}

\bibitem[\protect\citeauthoryear{{Henney} \& {Arthur}}{{Henney} \&
  {Arthur}}{1998}]{1998AJ....116..322H}
{Henney} W.~J.,  {Arthur} S.~J.,  1998, \mn@doi [\aj] {10.1086/300433}, \href
  {https://ui.adsabs.harvard.edu/abs/1998AJ....116..322H} {116, 322}

\bibitem[\protect\citeauthoryear{{Henney} \& {O'Dell}}{{Henney} \&
  {O'Dell}}{1999}]{1999AJ....118.2350H}
{Henney} W.~J.,  {O'Dell} C.~R.,  1999, \mn@doi [\aj] {10.1086/301087}, \href
  {https://ui.adsabs.harvard.edu/abs/1999AJ....118.2350H} {118, 2350}

\bibitem[\protect\citeauthoryear{{Hildebrand}}{{Hildebrand}}{1983}]{1983QJRAS..24..267H}
{Hildebrand} R.~H.,  1983, QJRAS, \href
  {http://adsabs.harvard.edu/abs/1983QJRAS..24..267H} {24, 267}

\bibitem[\protect\citeauthoryear{{Kim}, {Clarke}, {Fang}  \& {Facchini}}{{Kim}
  et~al.}{2016}]{2016ApJ...826L..15K}
{Kim} J.~S.,  {Clarke} C.~J.,  {Fang} M.,   {Facchini} S.,  2016, \mn@doi
  [\apjl] {10.3847/2041-8205/826/1/L15}, \href
  {https://ui.adsabs.harvard.edu/abs/2016ApJ...826L..15K} {826, L15}

\bibitem[\protect\citeauthoryear{{Kruijssen}, {Longmore}  \&
  {Chevance}}{{Kruijssen} et~al.}{2020}]{2020ApJ...905L..18K}
{Kruijssen} J.~M.~D.,  {Longmore} S.~N.,   {Chevance} M.,  2020, \mn@doi
  [\apjl] {10.3847/2041-8213/abccc3}, \href
  {https://ui.adsabs.harvard.edu/abs/2020ApJ...905L..18K} {905, L18}

\bibitem[\protect\citeauthoryear{{Krumholz}, {McKee}  \&
  {Bland-Hawthorn}}{{Krumholz} et~al.}{2019}]{2019ARA&A..57..227K}
{Krumholz} M.~R.,  {McKee} C.~F.,   {Bland-Hawthorn} J.,  2019, \mn@doi [\araa]
  {10.1146/annurev-astro-091918-104430}, \href
  {https://ui.adsabs.harvard.edu/abs/2019ARA&A..57..227K} {57, 227}

\bibitem[\protect\citeauthoryear{{Kurtovic} et~al.,}{{Kurtovic}
  et~al.}{2018}]{2018ApJ...869L..44K}
{Kurtovic} N.~T.,  et~al., 2018, \mn@doi [\apjl] {10.3847/2041-8213/aaf746},
  \href {https://ui.adsabs.harvard.edu/abs/2018ApJ...869L..44K} {869, L44}

\bibitem[\protect\citeauthoryear{{Lada} \& {Lada}}{{Lada} \&
  {Lada}}{2003}]{2003ARA&A..41...57L}
{Lada} C.~J.,  {Lada} E.~A.,  2003, \mn@doi [\araa]
  {10.1146/annurev.astro.41.011802.094844}, \href
  {https://ui.adsabs.harvard.edu/abs/2003ARA&A..41...57L} {41, 57}

\bibitem[\protect\citeauthoryear{{Lichtenberg}, {Golabek}, {Burn}, {Meyer},
  {Alibert}, {Gerya}  \& {Mordasini}}{{Lichtenberg}
  et~al.}{2019}]{2019NatAs...3..307L}
{Lichtenberg} T.,  {Golabek} G.~J.,  {Burn} R.,  {Meyer} M.~R.,  {Alibert} Y.,
  {Gerya} T.~V.,   {Mordasini} C.,  2019, \mn@doi [Nature Astronomy]
  {10.1038/s41550-018-0688-5}, \href
  {https://ui.adsabs.harvard.edu/abs/2019NatAs...3..307L} {3, 307}

\bibitem[\protect\citeauthoryear{{Longmore}, {Chevance}  \&
  {Kruijssen}}{{Longmore} et~al.}{2021}]{2021arXiv210301974L}
{Longmore} S.~N.,  {Chevance} M.,   {Kruijssen} J.~M.~D.,  2021, arXiv
  e-prints, \href {https://ui.adsabs.harvard.edu/abs/2021arXiv210301974L} {p.
  arXiv:2103.01974}

\bibitem[\protect\citeauthoryear{{Lucy}}{{Lucy}}{1999}]{1999A&A...344..282L}
{Lucy} L.~B.,  1999, A\&A, \href
  {http://adsabs.harvard.edu/abs/1999A%26A...344..282L} {344, 282}

\bibitem[\protect\citeauthoryear{{Madau} \& {Dickinson}}{{Madau} \&
  {Dickinson}}{2014}]{2014ARA&A..52..415M}
{Madau} P.,  {Dickinson} M.,  2014, \mn@doi [\araa]
  {10.1146/annurev-astro-081811-125615}, \href
  {https://ui.adsabs.harvard.edu/abs/2014ARA&A..52..415M} {52, 415}

\bibitem[\protect\citeauthoryear{{Mann} et~al.,}{{Mann}
  et~al.}{2014}]{2014ApJ...784...82M}
{Mann} R.~K.,  et~al., 2014, \mn@doi [\apj] {10.1088/0004-637X/784/1/82}, \href
  {https://ui.adsabs.harvard.edu/abs/2014ApJ...784...82M} {784, 82}

\bibitem[\protect\citeauthoryear{{Miley}, {Pani{\'c}}, {Booth}, {Ilee}, {Ida}
  \& {Kunitomo}}{{Miley} et~al.}{2021}]{2021MNRAS.500.4658M}
{Miley} J.~M.,  {Pani{\'c}} O.,  {Booth} R.~A.,  {Ilee} J.~D.,  {Ida} S.,
  {Kunitomo} M.,  2021, \mn@doi [\mnras] {10.1093/mnras/staa3517}, \href
  {https://ui.adsabs.harvard.edu/abs/2021MNRAS.500.4658M} {500, 4658}

\bibitem[\protect\citeauthoryear{{Miotello}, {Robberto}, {Potenza}  \&
  {Ricci}}{{Miotello} et~al.}{2012}]{2012ApJ...757...78M}
{Miotello} A.,  {Robberto} M.,  {Potenza} M. A.~C.,   {Ricci} L.,  2012,
  \mn@doi [\apj] {10.1088/0004-637X/757/1/78}, \href
  {https://ui.adsabs.harvard.edu/abs/2012ApJ...757...78M} {757, 78}

\bibitem[\protect\citeauthoryear{{Miyake} \& {Nakagawa}}{{Miyake} \&
  {Nakagawa}}{1993}]{1993Icar..106...20M}
{Miyake} K.,  {Nakagawa} Y.,  1993, \mn@doi [\icarus] {10.1006/icar.1993.1156},
  \href {https://ui.adsabs.harvard.edu/abs/1993Icar..106...20M} {106, 20}

\bibitem[\protect\citeauthoryear{{Natta}, {Testi}, {Neri}, {Shepherd}  \&
  {Wilner}}{{Natta} et~al.}{2004}]{2004A&A...416..179N}
{Natta} A.,  {Testi} L.,  {Neri} R.,  {Shepherd} D.~S.,   {Wilner} D.~J.,
  2004, \mn@doi [\aap] {10.1051/0004-6361:20035620}, \href
  {https://ui.adsabs.harvard.edu/abs/2004A&A...416..179N} {416, 179}

\bibitem[\protect\citeauthoryear{{Ndugu}, {Bitsch}  \& {Jurua}}{{Ndugu}
  et~al.}{2018}]{2018MNRAS.474..886N}
{Ndugu} N.,  {Bitsch} B.,   {Jurua} E.,  2018, \mn@doi [\mnras]
  {10.1093/mnras/stx2815}, \href
  {https://ui.adsabs.harvard.edu/abs/2018MNRAS.474..886N} {474, 886}

\bibitem[\protect\citeauthoryear{{Nicholson}, {Parker}, {Church}, {Davies},
  {Fearon}  \& {Walton}}{{Nicholson} et~al.}{2019}]{2019MNRAS.485.4893N}
{Nicholson} R.~B.,  {Parker} R.~J.,  {Church} R.~P.,  {Davies} M.~B.,  {Fearon}
  N.~M.,   {Walton} S. R.~J.,  2019, \mn@doi [\mnras] {10.1093/mnras/stz606},
  \href {https://ui.adsabs.harvard.edu/abs/2019MNRAS.485.4893N} {485, 4893}

\bibitem[\protect\citeauthoryear{{{\"O}berg}, {Murray-Clay}  \&
  {Bergin}}{{{\"O}berg} et~al.}{2011}]{2011ApJ...743L..16O}
{{\"O}berg} K.~I.,  {Murray-Clay} R.,   {Bergin} E.~A.,  2011, \mn@doi [\apjl]
  {10.1088/2041-8205/743/1/L16}, \href
  {https://ui.adsabs.harvard.edu/abs/2011ApJ...743L..16O} {743, L16}

\bibitem[\protect\citeauthoryear{{O'dell} \& {Wen}}{{O'dell} \&
  {Wen}}{1994}]{1994ApJ...436..194O}
{O'dell} C.~R.,  {Wen} Z.,  1994, \mn@doi [\apj] {10.1086/174892}, \href
  {https://ui.adsabs.harvard.edu/abs/1994ApJ...436..194O} {436, 194}

\bibitem[\protect\citeauthoryear{{Ormel}, {Liu}  \& {Schoonenberg}}{{Ormel}
  et~al.}{2017}]{2017A&A...604A...1O}
{Ormel} C.~W.,  {Liu} B.,   {Schoonenberg} D.,  2017, \mn@doi [\aap]
  {10.1051/0004-6361/201730826}, \href
  {https://ui.adsabs.harvard.edu/abs/2017A&A...604A...1O} {604, A1}

\bibitem[\protect\citeauthoryear{{Owen}}{{Owen}}{2020}]{2020MNRAS.495.3160O}
{Owen} J.~E.,  2020, \mn@doi [\mnras] {10.1093/mnras/staa1309}, \href
  {https://ui.adsabs.harvard.edu/abs/2020MNRAS.495.3160O} {495, 3160}

\bibitem[\protect\citeauthoryear{{Paxton}, {Bildsten}, {Dotter}, {Herwig},
  {Lesaffre}  \& {Timmes}}{{Paxton} et~al.}{2011}]{2011ApJS..192....3P}
{Paxton} B.,  {Bildsten} L.,  {Dotter} A.,  {Herwig} F.,  {Lesaffre} P.,
  {Timmes} F.,  2011, \mn@doi [\apjs] {10.1088/0067-0049/192/1/3}, \href
  {https://ui.adsabs.harvard.edu/abs/2011ApJS..192....3P} {192, 3}

\bibitem[\protect\citeauthoryear{{Paxton} et~al.,}{{Paxton}
  et~al.}{2013}]{2013ApJS..208....4P}
{Paxton} B.,  et~al., 2013, \mn@doi [\apjs] {10.1088/0067-0049/208/1/4}, \href
  {https://ui.adsabs.harvard.edu/abs/2013ApJS..208....4P} {208, 4}

\bibitem[\protect\citeauthoryear{{Paxton} et~al.,}{{Paxton}
  et~al.}{2015}]{2015ApJS..220...15P}
{Paxton} B.,  et~al., 2015, \mn@doi [\apjs] {10.1088/0067-0049/220/1/15}, \href
  {https://ui.adsabs.harvard.edu/abs/2015ApJS..220...15P} {220, 15}

\bibitem[\protect\citeauthoryear{{Reiter}}{{Reiter}}{2020}]{2020A&A...644L...1R}
{Reiter} M.,  2020, \mn@doi [\aap] {10.1051/0004-6361/202039334}, \href
  {https://ui.adsabs.harvard.edu/abs/2020A&A...644L...1R} {644, L1}

\bibitem[\protect\citeauthoryear{{Ricci}, {Testi}, {Natta}, {Neri}, {Cabrit}
  \& {Herczeg}}{{Ricci} et~al.}{2010}]{2010A&A...512A..15R}
{Ricci} L.,  {Testi} L.,  {Natta} A.,  {Neri} R.,  {Cabrit} S.,   {Herczeg}
  G.~J.,  2010, \mn@doi [\aap] {10.1051/0004-6361/200913403}, \href
  {https://ui.adsabs.harvard.edu/abs/2010A&A...512A..15R} {512, A15}

\bibitem[\protect\citeauthoryear{{Richling} \& {Yorke}}{{Richling} \&
  {Yorke}}{2000}]{2000ApJ...539..258R}
{Richling} S.,  {Yorke} H.~W.,  2000, \mn@doi [\apj] {10.1086/309198}, \href
  {https://ui.adsabs.harvard.edu/abs/2000ApJ...539..258R} {539, 258}

\bibitem[\protect\citeauthoryear{{Robberto}, {Beckwith}  \&
  {Panagia}}{{Robberto} et~al.}{2002}]{2002ApJ...578..897R}
{Robberto} M.,  {Beckwith} S.~V.~W.,   {Panagia} N.,  2002, \mn@doi [\apj]
  {10.1086/342615}, \href
  {https://ui.adsabs.harvard.edu/abs/2002ApJ...578..897R} {578, 897}

\bibitem[\protect\citeauthoryear{{Rodriguez} et~al.,}{{Rodriguez}
  et~al.}{2018}]{2018ApJ...859..150R}
{Rodriguez} J.~E.,  et~al., 2018, \mn@doi [\apj] {10.3847/1538-4357/aac08f},
  \href {https://ui.adsabs.harvard.edu/abs/2018ApJ...859..150R} {859, 150}

\bibitem[\protect\citeauthoryear{{Scally} \& {Clarke}}{{Scally} \&
  {Clarke}}{2001}]{2001MNRAS.325..449S}
{Scally} A.,  {Clarke} C.,  2001, \mn@doi [\mnras]
  {10.1046/j.1365-8711.2001.04274.x}, \href
  {https://ui.adsabs.harvard.edu/abs/2001MNRAS.325..449S} {325, 449}

\bibitem[\protect\citeauthoryear{{Schoonenberg}, {Liu}, {Ormel}  \&
  {Dorn}}{{Schoonenberg} et~al.}{2019}]{2019A&A...627A.149S}
{Schoonenberg} D.,  {Liu} B.,  {Ormel} C.~W.,   {Dorn} C.,  2019, \mn@doi
  [\aap] {10.1051/0004-6361/201935607}, \href
  {https://ui.adsabs.harvard.edu/abs/2019A&A...627A.149S} {627, A149}

\bibitem[\protect\citeauthoryear{{Sellek}, {Booth}  \& {Clarke}}{{Sellek}
  et~al.}{2020}]{2020MNRAS.492.1279S}
{Sellek} A.~D.,  {Booth} R.~A.,   {Clarke} C.~J.,  2020, \mn@doi [\mnras]
  {10.1093/mnras/stz3528}, \href
  {https://ui.adsabs.harvard.edu/abs/2020MNRAS.492.1279S} {492, 1279}

\bibitem[\protect\citeauthoryear{{Siess}, {Dufour}  \& {Forestini}}{{Siess}
  et~al.}{2000}]{2000A&A...358..593S}
{Siess} L.,  {Dufour} E.,   {Forestini} M.,  2000, \aap, \href
  {https://ui.adsabs.harvard.edu/abs/2000A&A...358..593S} {358, 593}

\bibitem[\protect\citeauthoryear{{Tazzari} et~al.,}{{Tazzari}
  et~al.}{2017}]{2017A&A...606A..88T}
{Tazzari} M.,  et~al., 2017, \mn@doi [\aap] {10.1051/0004-6361/201730890},
  \href {https://ui.adsabs.harvard.edu/abs/2017A&A...606A..88T} {606, A88}

\bibitem[\protect\citeauthoryear{{Tsamis}, {Flores-Fajardo}, {Henney}, {Walsh}
  \& {Mesa-Delgado}}{{Tsamis} et~al.}{2013}]{2013MNRAS.430.3406T}
{Tsamis} Y.~G.,  {Flores-Fajardo} N.,  {Henney} W.~J.,  {Walsh} J.~R.,
  {Mesa-Delgado} A.,  2013, \mn@doi [\mnras] {10.1093/mnras/stt145}, \href
  {https://ui.adsabs.harvard.edu/abs/2013MNRAS.430.3406T} {430, 3406}

\bibitem[\protect\citeauthoryear{{Walsh}, {Millar}  \& {Nomura}}{{Walsh}
  et~al.}{2013}]{2013ApJ...766L..23W}
{Walsh} C.,  {Millar} T.~J.,   {Nomura} H.,  2013, \mn@doi [\apjl]
  {10.1088/2041-8205/766/2/L23}, \href
  {https://ui.adsabs.harvard.edu/abs/2013ApJ...766L..23W} {766, L23}

\bibitem[\protect\citeauthoryear{{Winter}, {Clarke}, {Rosotti}, {Ih},
  {Facchini}  \& {Haworth}}{{Winter} et~al.}{2018}]{2018MNRAS.478.2700W}
{Winter} A.~J.,  {Clarke} C.~J.,  {Rosotti} G.,  {Ih} J.,  {Facchini} S.,
  {Haworth} T.~J.,  2018, \mn@doi [\mnras] {10.1093/mnras/sty984}, \href
  {https://ui.adsabs.harvard.edu/abs/2018MNRAS.478.2700W} {478, 2700}

\bibitem[\protect\citeauthoryear{{Winter}, {Clarke}, {Rosotti}, {Hacar}  \&
  {Alexander}}{{Winter} et~al.}{2019}]{2019MNRAS.490.5478W}
{Winter} A.~J.,  {Clarke} C.~J.,  {Rosotti} G.~P.,  {Hacar} A.,   {Alexander}
  R.,  2019, \mn@doi [\mnras] {10.1093/mnras/stz2545}, \href
  {https://ui.adsabs.harvard.edu/abs/2019MNRAS.490.5478W} {490, 5478}

\bibitem[\protect\citeauthoryear{{Winter}, {Kruijssen}, {Longmore}  \&
  {Chevance}}{{Winter} et~al.}{2020}]{2020Natur.586..528W}
{Winter} A.~J.,  {Kruijssen} J.~M.~D.,  {Longmore} S.~N.,   {Chevance} M.,
  2020, \mn@doi [\nat] {10.1038/s41586-020-2800-0}, \href
  {https://ui.adsabs.harvard.edu/abs/2020Natur.586..528W} {586, 528}

\bibitem[\protect\citeauthoryear{{van Terwisga} et~al.,}{{van Terwisga}
  et~al.}{2020}]{2020A&A...640A..27V}
{van Terwisga} S.~E.,  et~al., 2020, \mn@doi [\aap]
  {10.1051/0004-6361/201937403}, \href
  {https://ui.adsabs.harvard.edu/abs/2020A&A...640A..27V} {640, A27}

\makeatother
\end{thebibliography}

\label{lastpage}

\end{document}